\documentclass[12pt,a4paper]{article}
\usepackage{amsmath}
\usepackage[font={footnotesize,it}]{caption}
\usepackage[top=1.5in, bottom=1.5in, left=1.1in, right=1.1in]{geometry}

\DeclareCaptionStyle{italic}[justification=centering]{labelfont={bf},textfont={it},labelsep=colon}
\usepackage{graphicx,psfrag,epsf}
\usepackage{enumerate}
\usepackage{natbib}
\usepackage{url, setspace}
\usepackage{booktabs, subfig, bm, paralist,mathpazo,tikz,longtable,microtype,dsfont,rotating} 
\usepackage[pdftex,colorlinks=true]{hyperref}
\usepackage{mathtools}	      
\usepackage{appendix}
\usepackage{mathrsfs}
\usepackage{multirow,orcidlink}
\usepackage{xurl}
\usepackage{amssymb}
\usepackage{prettyref}
\newrefformat{hypo}{H\ref{#1}}
\newcounter{hypothesis}
\renewcommand{\thehypothesis}{\arabic{hypothesis}}
\definecolor{darkblue}{rgb}{0,0,.6}
\hypersetup{citecolor=darkblue,linkcolor=darkblue,urlcolor=darkblue}

\newtheorem{thm}{\sc Theorem}

\newtheorem{prop}{\sc Proposition}

\newtheorem{rem}{\sc Remark}
\newcommand{\xdownarrow}[1]{%
  {\left\downarrow\vbox to #1{}\right.\kern-\nulldelimiterspace}
}
\newcommand{\blind}{0}

\newcommand{\X}{\mathcal{X}}
\newcommand{\E}{\text{E}}
\newcommand{\var}{\text{Var}}
\newcommand{\cov}{\text{Cov}}
\newcommand{\corr}{\text{Corr}}

\newcommand{\doublehat}[1]{\,\widehat{\mkern-5mu\widehat{#1}}}
\graphicspath{{plots/}}

\newsavebox\CBox
\def\textBF#1{\sbox\CBox{#1}\resizebox{\wd\CBox}{\ht\CBox}{\textbf{#1}}}

\def\R{\mathbb{R}}
\def\Z{\mathbb{Z}}
\def\bX{\boldsymbol{X}}
\date{}
\newcommand{\Rlogo}{\protect\includegraphics[height=1.8ex,keepaspectratio]{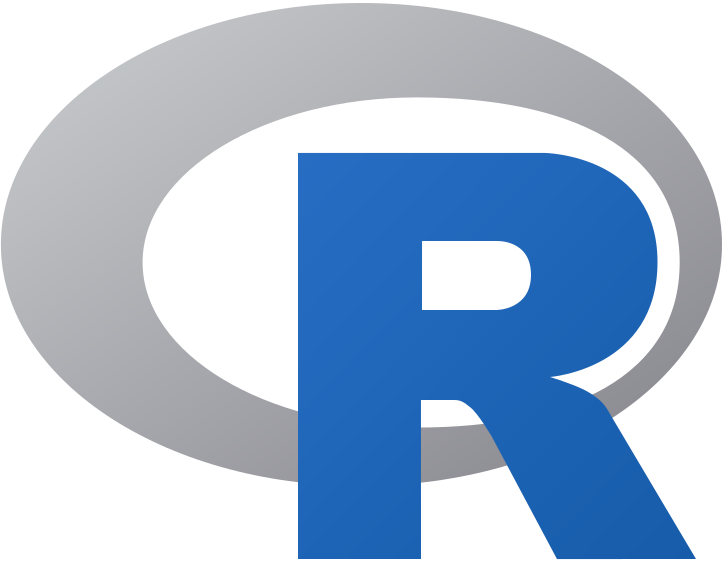}}

\begin{document}

\def\spacingset#1{\renewcommand{\baselinestretch}
{#1}\small\normalsize} \spacingset{1}

\if0\blind
{
  \title{\bf Density-valued time series: Nonparametric density-on-density regression}
\author{\normalsize Fr\'ed\'eric Ferraty \orcidlink{0000-0002-5537-2667}
 \hspace{.2cm}\\
\normalsize Toulouse Mathematics Institute\\ 
\normalsize University of Toulouse\\ 
\\
\normalsize  Han Lin Shang \orcidlink{0000-0003-1769-6430}
  \hspace{.2cm}\\
\normalsize     Department of Actuarial Studies and Business Analytics \\
\normalsize     Macquarie University}
  \maketitle
} \fi

\if1\blind
{
   \title{\bf Density-valued time series: Nonparametric density-on-density regression}
   \author{}
   \maketitle
} \fi

\bigskip

\begin{abstract}
This paper is concerned with forecasting probability density functions. Density functions are nonnegative and have a constrained integral; thus, they do not constitute a vector space. Implementing unconstrained functional time-series forecasting methods is problematic for such nonlinear and constrained data. A novel forecasting method is developed based on a nonparametric function-on-function regression, where both the response and the predictor are probability density functions. Asymptotic properties of our nonparametric regression estimator are established, as well as its finite-sample performance through a series of Monte-Carlo simulation studies. Using COVID-19 data from the French department and age-specific period life tables from the United States, we assess and compare the finite-sample forecast accuracy of the proposed method with several existing methods.
\\

\noindent \textit{Short Run Title}: Nonparametric density-on-density regression 
\end{abstract}

\newpage
\spacingset{1.6}

\section{Introduction}

Recent advances in computer storage and data collection have enabled researchers to record data of characteristics varying over a continuum (time, space, depth, wavelength, etc.). In various branches of science, data are collected using a spectrometer, rain gauges, electroencephalographs, or even a high-performance computer. In all of these cases, a number of subjects are observed densely over time, space, or both. Through the application of interpolation or smoothing techniques, these data become functions that can be represented as a curve, image, or shape. Functional data analysis has arisen as a field of statistics that provides statistical tools for analysing this type of information. For an overview of functional data analysis, \cite{RS05} presented several state-of-the-art statistical techniques, while \cite{FV06} listed a range of nonparametric techniques. \cite{HE15} provided theoretical foundations with an introduction to linear operators, while \cite{Bosq00} studied the functional autoregressive process from a linear operator perspective. An introduction to temporally dependent functional data can be found in \cite{KR17}, while \cite{MG22} studied spatially-dependent functional data.

As an integral part of functional data analysis and time series analysis, functional time series consist of random functions observed at a time interval. Functional time series can be classified into two categories depending on whether the continuum is also a time variable. On the one hand, when the continuum is not a time variable, functional time series can also arise when observations in a period can be considered as finite-dimensional realisations of an underlying continuous function. Examples include yearly age-specific mortality rates \citep[see, e.g.,][]{CM09, HS09} and a time series of near-infrared spectroscopy curves \citep[see, e.g.,][]{SCS22}. On the other hand, functional time series can arise from measurements obtained by separating a continuous-time stochastic process into natural consecutive intervals, for example, days, weeks, or years \citep[see, e.g.,][]{Bosq00, BCS00, AS03, APS06, FV06, HK12, KRS17, Shang17}. 

Functional data that are samples of one-dimensional random probability density functions (PDFs) are common. Examples include income distributions \citep{KU01}, distributions of times when bids are submitted in an online auction \citep{JSP+08}, functional connectivity in the brain \citep{PM19}, distribution of image features from head CT scans \citep{SNM+20}, distributions of stock returns \citep{HLZ16, BG19, ZKP21}, the age distribution of fertility rates \citep{MS15}, and age distribution of life-table death counts in demography \citep{BCO+17}. 

To model density-valued functional data, \cite{JR92} estimated the density function by a kernel density estimator and displayed the density function via principal components. \cite{NG07} proposed regression trees where the response variable is PDF. \cite{vanderLinde08} introduced Bayesian functional principal component analysis and applied it to simulated data consisting of nonparametric density estimates. \cite{SKJ+11} proposed a time-warping function where the square root transformation of densities resides in Hilbert space. \cite{Dai22} considered statistical inference of the Fr\'{e}chet mean in the Hilbert sphere with application to random densities. \cite{ZKP21} proposed Wasserstein autoregressive models for density time series. \cite{SH20, SH25b, SH25} considered the centred log-ratio transformation, 

While any PDF $f$ may be thought of as a random element of a Hilbert space, it does not reside in linear space due to summability and non-negativity constraints (i.e. $f \geq 0$ and $\int f = 1$); if $f$, $g$ are two PDFs and $a$, $b$ two real values, then the combination $a \, f \, + \, b \, g$ is no longer a PDF (except in the special case $b = 1-a$ with $a\geq 0$). Consequently, the standard space of square-integrable functions $\mathcal{L}^2$ is not the most appropriate for representing PDFs. A natural way to deal with such constraints is to peel them away by an invertible transformation that maps densities onto a linear space. From an extrinsic viewpoint, such transformations include the Bayes Hilbert space approach \citep[see, e.g.,][]{HMTH16, MSG22}, compositional data analysis (CoDa) \citep{SH20}, cumulative distribution function transformation \citep{SH25b}, $\alpha$ transformation \citep{SH25} and log quantile density transformation (LQDT) \citep[see, e.g.,][]{PM16, HMP19}. \cite{KMP+19} compared the finite-sample performance of density estimation and forecasting among the LQDT, CoDa, unconstrained functional principal component regression of \cite{HZ18}, and a skewed-$t$ distribution of \cite{Wang12}. \cite{KMP+19} recommended the LQDT and CoDa, two benchmark methods in our empirical studies.

\begin{figure}[!htbp]
\centering
\includegraphics[width = 8.7cm]{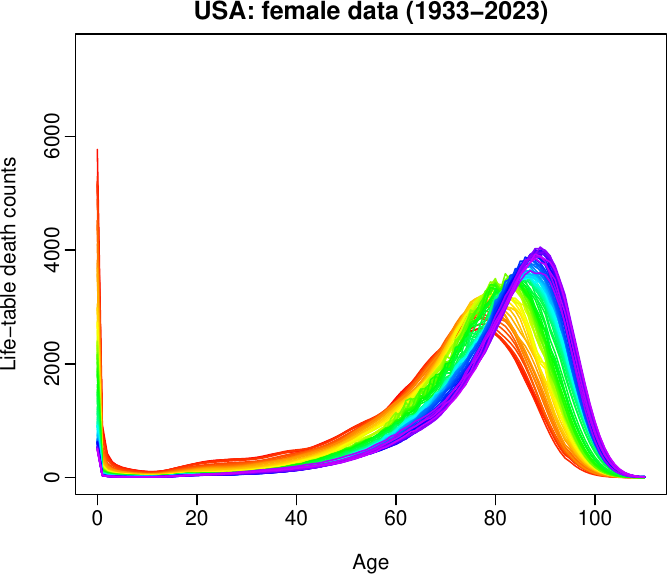}
\quad
\includegraphics[width = 8.7cm]{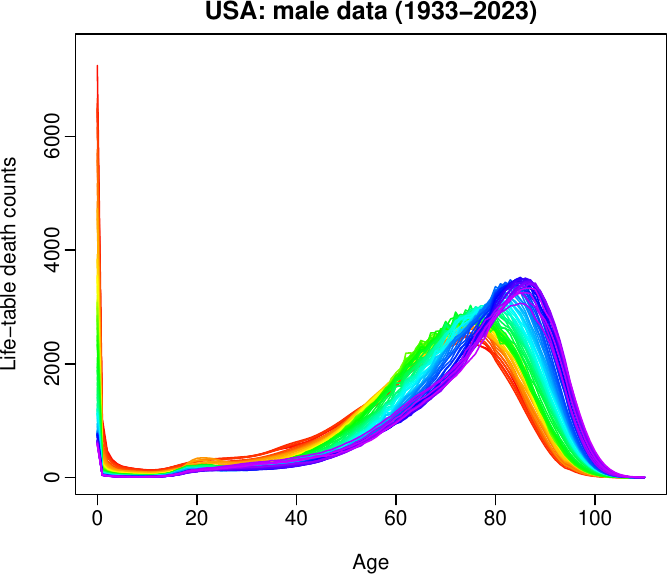}
\caption{\small{Rainbow plots of the age distribution of life-table death count from 1933 to 2023 in a single-year group. The life-table radix is 100,000 for each year. The life-table death counts in the oldest years are shown in red, while the most recent years are in violet. Curves are ordered chronologically according to the colours of the rainbow.}}\label{fig:1}
\end{figure}

In this paper, we focus on a random process that takes values in a space of densities. We aim to forecast the future density nonparametrically, given those observed in the past. Density forecasting is of great importance, especially in demography. In the realm of actuarial statistics, an important use of density forecasting is to predict the age distribution of life-table death counts as a means of computing survival probabilities. In such a model, any incorrect specification of the forecast density can produce an inaccurate estimate of the value-at-risk and make the asset holder, i.e., insurance companies, unable to control risk \citep[see, e.g.,][]{Shang13}. In Figure~\ref{fig:1}, we present a time series of age-specific life-table death counts for the United States of America (USA) from 1933 to 2023 obtained from the \cite{HMD22}.

This setting faces three non-standard frameworks: 
\begin{inparaenum}
\item[1)] nonparametric density-on-density regression, 
\item[2)] the dependence of the sequence of observed density functions, and 
\item[3)] the convexity property of the space of densities that do not form a linear space. 
\end{inparaenum}
To take into account the nonlinear feature of the space of PDFs, we consider the Bayes Hilbert space approach \citep[see, e.g.,][]{EgozDP06, VanBEP10, VanBEP14, HMTH16}. The Bayes Hilbert spaces provide a mathematical environment that allows the extension of Aitchison geometry from finite-dimensional compositional data \citep{Aitchison86, PawlETD15} to an infinite-dimensional setting. The Bayes Hilbert space is a linear space of PDFs designed for statistical models involving random PDFs \citep[see, e.g.,][]{Delicado11, HMTH16, SB19, JeonP20}).

A nonparametric density-on-density regression, where the response is one-step-ahead of the predictor in a time series of densities, is defined in the Bayes Hilbert space. A distance-based approach with a kernel-type estimator is proposed. Its flexibility makes the implementation attractive, since various distance metrics may be considered. Its simplicity also makes the implementation easy and computationally fast as it does in the more standard nonparametric function-on-function regression setting (see, e.g., \citealp{FVV12, Lian12}). The advantage of nonparametric modelling is to capture the possible nonlinear dependence between two consecutive random PDFs. Unfortunately, instead of observing the density-valued random process, we sometimes only have samples drawn from these density functions \citep[see also][]{KMP+19, SH20, ZKP21}. This is why our methodology often includes, in its first step, the estimation of the densities before nonparametrically estimating the relationship between densities.

The paper is organised as follows. Section~\ref{sec:2.0} reminds of the basic geometric concepts of the Bayes Hilbert space. In Section~\ref{sec:2.1}, we describe a model, estimation and forecasting scheme for our nonparametric density-on-density regression. In Section~\ref{sec:2.2}, we present asymptotic properties associated with our Bayes Nadaraya-Watson (NW) estimator. In particular, we establish a condition that ensures the existence of a stationary PDF-valued random process consistent with the proposed nonparametric PDF-on-PDF regression model. We also establish the asymptotic behaviour of the mean integrated squared error of the kernel density estimator when the target is a PDF-valued random variable.  In Section~\ref{sec:3.1}, we describe how we implement the kernel density estimator, the selection of optimal bandwidth, and the criterion used to assess density estimation accuracy. In Section~\ref{sec:3.3}, we evaluate and compare our Bayes NW estimator in the nonparametric density-on-density regression via a series of simulation studies. A French COVID-19 hospitalisation data analysis is shown in Section~\ref{sec:3.6}, and a data analysis of age-specific life-table death count in the United States (U.S.) is presented in Section~\ref{sec:3.5}. The conclusion is given in Section~\ref{sec:5}, along with some ideas on how the methodology can be further extended.

\section{Nonparametric density-on-density regression}\label{sec:2}

We first briefly remind the definition of the Bayes Hilbert space before introducing the density-on-density regression model. It is essential to understand how this vector space of PDFs works and its associated geometry. 

\subsection{Bayes Hilbert space: a vector space of PDFs}\label{sec:2.0}

\paragraph{Bayes Hilbert space.} Let $\mathcal{B}_2(I)$ be the set of bounded PDFs $f$ with continuous support $I=[a,\, b]$ for which the natural logarithm is square integrable: 
\begin{equation*}
\mathcal{B}_2(I)= \left\{ f:I \mapsto ]c, \, +\infty[ \mbox{ with }c>0, \, \int_I f = 1, \, \int_I \{\ln(f)\}^2 \,  < \, +\infty \right\}. 
\end{equation*}
Remark: to simplify notation, we omit the arguments of the integrand whenever no ambiguity arises. Two fundamental operators $\oplus$ and $\odot$ respectively called {\em perturbation} and {\em powering} are defined for any $f$, $g$ in $\mathcal{B}_2(I)$ and $r$ in $\R$: 
\begin{equation*}
\displaystyle (f \oplus g)(t) = \frac{f(t) \, g(t)}{ \int_I f(u) \, g(u) \, du}, \qquad \displaystyle (r \odot f)(t) = \frac{f^r(t)}{ \int_I f^r(u) \, du}. 
\end{equation*}
The perturbation is commutative and associative: 
\begin{equation*}
f \oplus g = g \oplus f, \qquad (f \oplus g) \oplus h = f \oplus (g \oplus h). 
\end{equation*}
Let $\displaystyle {\bf 0}_\mathcal{B}=\frac{1}{b-a}1_I$, the uniform PDF defined on $I$, be the neutral element ($f \oplus {\bf 0}_\mathcal{B} = {\bf 0}_\mathcal{B} \oplus f = f$) for the perturbation and $f^{-1}$ be the unique element such that $f \oplus f^{-1} = f^{-1} \oplus f = {\bf 0}_\mathcal{B}$. The power is associative ($r \odot (s \odot f) = (r \, s) \odot f$), distributive with respect to the perturbation ($r \odot (f \oplus g) = (r \odot f) \oplus (r \odot g))$ and the scalar addition ($(r + s) \odot f = (r \odot f) \oplus (s \odot f)$), and 1 is its neutral element. Based on perturbation and powering, a third operator $\ominus$ called {\em perturbation-substraction} is defined: 
\begin{equation*}
f \ominus g = f \oplus (-1) \odot g. 
\end{equation*}
In $\mathcal{B}_2(I)$, the inner product is defined as: 
\begin{equation*}
\displaystyle \langle f , g \rangle_\mathcal{B} = \frac{1}{2 \, (b-a)}\int_I \int_I \ln \frac{f(u)}{f(v)} \, \ln \frac{g(u)}{g(v)}  \, du \, dv 
\end{equation*}
and one has useful properties:  
\begin{inparaenum}
\item[1)] $(\mathcal{B}_2(I), \, \oplus,\, \odot, \,  \langle \cdot , \cdot \rangle_\mathcal{B})$ is a vector space, 
\item[2)] $(\mathcal{B}_2(I), \, \oplus,\, \odot, \,  \langle \cdot , \cdot \rangle_\mathcal{B})$ forms a separable Hilbert space called Bayes Hilbert space \citep[see][]{EgozDP06, VanBEP14}. 
\end{inparaenum}
Finally, for any $f$, $g$ in $\mathcal{B}_2(I)$, $\|f\|_\mathcal{B}=\sqrt{\langle f, f \rangle_\mathcal{B}}$ is the norm associated with the inner product. One can remark that $ \| f \ominus g \|_\mathcal{B} = 0$ is equivalent to $f \ominus g = {\bf 0}_\mathcal{B}$, which results in $f(u) = g(u)$ for any $u \in  I$. 

\paragraph{Isometry between $\mathcal{B}_2(I)$ and $\mathcal{L}^2_0(I)$.} Let  $\mathcal{L}^2_0(I)$ be the space of square-integrable real functions $f$ on $I$ with $\int_I f =0$. There exists an isometric isomorphism between $\mathcal{B}_2(I)$ and $\mathcal{L}^2_0(I)$ called the centred $\log$-ratio (clr) transformation, defined for any PDF $f$  in $\mathcal{B}_2(I)$ and any $u$ in $I$: 
\begin{equation*}
\text{clr}(f)(u) = \ln f(u) - (b-a)^{-1}\int_I \ln f. 
\end{equation*}
In addition of the isometry property of the clr transformation $\| f \ominus g \|_\mathcal{B} = \| \text{clr}(f) - \text{clr}(g) \|$ where $\| g \| = \int_I g^2$, $\text{clr}(f \oplus g) = \text{clr}(f) + \text{clr}(g)$, $\text{clr} (s \odot f) = s \times \text{clr}(f)$, and its inverse transformation 
\begin{equation*}
\text{clr}^{-1}(g) = \frac{\exp(g)}{\int_I \exp(g)},
\end{equation*} 
for any $g$ in  $\mathcal{L}^2_0(I)$. 

After recalling the Bayes Hilbert space framework, we introduce our nonparametric density-on-density regression for the analysis of time series of PDFs.


\subsection{Model, estimation, and sequential forecasting scheme}\label{sec:2.1}

\subsubsection{Density-on-density regression model} 

Let $\{f_t\}_{t\in \Z}$ be a stochastic process taking values in $\mathcal{B}_2(I)$ and consider the $N+1$ random PDFs $(f_1, \ldots, f_{N+1})$. The main goal is to forecast a future PDF $f_{N+2}$ given those observed in the past ($f_{N+1}, \, f_N, \ldots, f_1$). To this end, we propose the nonparametric density-on-density regression model
\begin{equation}
f_{t+1} = m(f_t)\oplus\epsilon_t, \qquad t=1,\dots,N, \label{eq:1}
\end{equation}
where $m(\cdot)$ is an unknown smooth operator mapping $\mathcal{B}_2(I)$ into $\mathcal{B}_2(I)$, and for each $t$, $\E(\epsilon_t | f_t)={\bf 0}_\mathcal{B}$, where the conditional expectation is defined for $\mathcal{B}_2(I)$-valued random PDFs. The definition of a conditional expectation for Hilbert-valued random variables can be found in \cite{Bosq00}. Under Model~\eqref{eq:1}, it is easy to check that, for each $t$, $\E(\epsilon_t | f_t)={\bf 0}_\mathcal{B}$ is equivalent to $m(f)= \E\left( f_{t+1} | f_t = f \right)$ (see Section~\ref{app:A} in the Appendix). Our main challenge is to define for any given density $f$ an estimator $\widetilde{m}(f)$ of the unknown regression operator $m(f)= \E\left( f_{t+1} | f_t = f \right)$ fulfilling density features (nonnegative functions that integrate to one) and good asymptotic behaviour while no reducing functional form (i.e. linearity, additivity, functional index model, etc) of the unknown regression operator $m(\cdot)$ is assumed.

\subsubsection{Kernel estimator} 

With some smoothness properties, the functional form of $m(\cdot)$ is often estimated in a data-driven manner. There are a number of nonparametric functional estimators, such as the functional Nadaraya-Watson (NW) estimator \citep{FV06}, the functional local linear estimator (\citealp{BEM11, FN22}), the functional $k$-nearest neighbour estimator (\citealp{BFV09}, for the scalar response and \citealp{Lian11}, for the functional response), the functional smoothing splines \citep{CG10}, the functional wavelet estimator \citep{AS03}, the distance-based local linear estimator \citep{BDF10} and the functional neural network \citep{RV06}. Throughout the paper, we introduce the functional Bayes NW estimator, i.e., the functional NW estimator in the Bayes-Hilbert space, because of its simplicity and mathematical elegance. The essence of functional Bayes NW kernel smoothing is to allow flexible estimation of the unknown regression operator. The functional Bayes NW estimator of the conditional mean can be defined as
\begin{equation} \label{def:regression_estimator}
\widetilde{m}_N(f) \, =  \, \bigoplus^N_{t=1} w_{h_{\text{reg}}}(f_t, \, f) \, \odot \, f_{t+1} \ \mbox{ with } \ w_{h_{\text{reg}}}(f_t, \, f) \, = \,\frac{K_{\text{reg}} ( h_{\text{reg}}^{-1} \, \| f_t \ominus f \|_\mathcal{B} ) }{\sum^N_{t=1}K_{\text{reg}} ( h_{\text{reg}}^{-1} \, \| f_t \ominus f \|_\mathcal{B} )},
\end{equation}
where $K_{\text{reg}}(\cdot)$ is an asymmetric kernel function and $h_{\text{reg}}$ is a positive smoothing parameter (i.e. bandwidth). The subscript {\em reg} in the notation refers to the estimator of the regression operator.  The vector space property of $\mathcal{B}_2(I)$ ensures that the Bayes NW estimator is a PDF. Depending on the kernel function and the bandwidth parameter, the nonparametric estimator in~\eqref{def:regression_estimator} is a weighted average of the observed densities and has two interesting properties: it fulfils the PDF constraints by construction, and due to its simplicity, its implementation is straightforward and computationally fast.

In practice, for each $t=1,\ldots,N+1$, we observe the $n$ real random variable (rrv) $X_{t,1},\ldots, X_{t, n}$ that shares the same marginal density $f_t$. In other words, PDFs are not directly observable. This situation is particularly common in epidemiology. For example, suppose that at each date~$t$, confirmed COVID-19 cases for all French departments are observed. In this case, $X_{t,j}$ is the COVID-19 confirmed cases at date $t$ of the department $j$, and the sample size $n$ is equal to the total number of departments. This is why our methodology first focusses on kernel density estimators $\widehat{f}_1, \ldots, \widehat{f}_{N+1}$ of the $N+1$ probability densities $f_1, \ldots, f_{N+1}$. In the second step, we just plug the estimated densities $\widehat{f}_1, \ldots, \widehat{f}_{N+1}$ into the nonparametric density-on-density regression estimator (\ref{def:regression_estimator}), which provides our definitive kernel estimation:
\begin{equation} \label{def:regression_estimator_practice}
\widehat{m}_N(f) \, =  \, \bigoplus^N_{t=1} w_{h_{\text{reg}}}(\widehat{f}_t, \, f) \, \odot \, \widehat{f}_{t+1}, 
\end{equation}
where $w_{h_{\text{reg}}}(\widehat{f}_t, \, f)$ is defined in (\ref{def:regression_estimator}). A standard by-product of this estimating procedure in the time-series context is the following one-step-ahead sequential forecasting scheme. From the estimated density functions $\widehat{f}_1, \ldots, \widehat{f}_{N+1}$, we compute $\doublehat{f}_{N+2} = \widehat{m}_N(\widehat{f}_{N+1})$ which is the prediction of the density $f_{N+2}$. To forecast the next density, just consider the sample of random densities completed with the predicted one (i.e. $\widehat{f}_1, \ldots, \widehat{f}_{N+1}, \doublehat{f}_{N+2}$) and build $\doublehat{f}_{N+3} = \widehat{m}_{N+1}(\doublehat{f}_{N+2})$. This one-step-ahead forecasting scheme can be iterated to obtain $\doublehat{f}_{N+H}$ at any reasonable forecast horizon~$H$.

\subsubsection{Probabilistic setting and dependence structure} \label{sec:Prob_and_dep}

To better understand the dependence structure of our data, our framework combines two embedded random mechanisms: 
\begin{inparaenum}
\item[1)] the random process $\{f\}_{t\in \Z}$ generating the random PDFs $\{f_1, \ldots, f_{N+1}\}$ and 
\item[2)] the random process providing for each date $t$ the $n$-tuple  $\bX_t = \left(X_{t,1}, \ldots, X_{t,n}\right)$ for estimating the non-observable random PDF $f_t$. 
\end{inparaenum}
The scheme hereafter sums up the overall random situation: 
\begin{center}
\begin{tabular}{rcclcccl}
Density-valued process &$f_1$ & $\longrightarrow$ & $f_2 = m(f_1) \oplus \epsilon_1$ &  $\longrightarrow$ & $\cdots$  &  $\longrightarrow$ & $f_{N+1} = m(f_N) \oplus \epsilon_N$ \\ 
 & $\big\downarrow$ & & $\big\downarrow$ & & & & $\big\downarrow$ \\
Observed data & $\bX_1$ &  &  $\bX_2$ &  &   &  &  $\bX_{N+1}$\\ 
 & $\big\downarrow$ & & $\big\downarrow$ & & & & $\big\downarrow$ \\
Estimated densities & $\widehat{f}_1$ & & $\widehat{f}_2$ & & & & $\widehat{f}_{N+1}$
\end{tabular}
\end{center}
{\em Probabilistic setting}. Let $(\Omega, \mathcal{A}, P)$ be a probability space; the PDF-valued random process $\{f_t\}_{t\in \Z}$ is such that, for all $\omega$ in $\Omega$, $f_t(\omega,.)$ is in the Bayes Hilbert space $\mathcal{B}_2(I)$ and for each $t$, the map $\omega \mapsto f_t(\omega,.)$ is a measurable mapping from $(\Omega, \mathcal{A}, P)$ to $(\mathcal{B}_2(I), \mathcal{G})$ with $\mathcal{G}$ a $\sigma$-algebra. For each $t$, we assume that 
\begin{enumerate}
\item[1)] the PDF of $\bX_t$ conditionally to $f_t$ is $f_t^n$, which means that, conditionally to $f_t$, $X_{t,1},\ldots, X_{t,n}$ are independently and identically distributed (iid) with PDF of $X_{t,1}$ equals to $f_t$, 
\item[2)] the vector random process $\{\bX_t\}_{t\in \Z}$ is independent across $t$ conditionally to the PDF-valued random process $\{f_t,\}_{t\in \Z}$, which means that, for any finite tuple $(t_1, \ldots, t_k)\in \Z^k$, the PDF of $(\bX_{t_1}, \ldots, \bX_{t_k})$ conditionally to $\{f_s\}_{s\in \Z}$ is equal to $\Pi_{j=1}^k\, f_{t_j}^n$.
\end{enumerate}

\noindent {\em Dependence structure}. The PDF-valued process $\{f_t\}_{t\in \Z}$ is assumed to be a stationary $\rho$-mixing process \citep[see, e.g.,][]{Rio93, Bradley05}. Set $\mathcal{F}_p^q = \sigma(f_t: \, p\leq t \leq q)$ the $\sigma$-algebra generated by $f_p,\ldots, f_q$ and let $\mathcal{L}^2(\mathcal{F}_p^q)$ denote the square-integrable, $\mathcal{F}_p^q$-measurable rrv space. The quantity 
\begin{equation} \label{def:mixing_coefficient}
\rho_f(H) = \sup\big\{ \left| \text{Corr}(U, \, V) \right|, \, U\in \mathcal{L}^2(\mathcal{F}_{-\infty}^0), \, V \in \mathcal{L}^2(\mathcal{F}_{H}^{+\infty})\big\}
\end{equation}
is called $\rho$-mixing coefficient of $\{f_t\}_{t\in\Z}$; it measures the dependence between $\mathcal{F}_{-\infty}^0$ and $\mathcal{F}_{H}^{+\infty}$ in terms of correlation. $\rho_f(H)$ is between 0 and 1; the closer this quantity is to 0, the weaker the dependence. By definition, the PDF-valued process $\{f_t\}_{t\in\Z}$ is $\rho$-mixing as soon as $\rho_f(H) \rightarrow 0$ as $H \rightarrow \infty$. The dependence structure is summarized in the following hypothesis:
\begin{itemize}
\item[\refstepcounter{hypothesis} \label{hypo:dependence} (H\thehypothesis)]
The random processes $\{f_t\}_{t\in \Z}$ and $\{\bX_t\}_{t\in \Z}$ are such that:
\begin{itemize}
\item[$\bullet$] $\{f_t\}_{t\in \Z}$ is a stationary $\rho$-mixing process,
\item[$\bullet$] for any finite tuple $(t_1, \ldots, t_k) \in \Z^k$, the PDF of $(\bX_{t_1}, \ldots, \bX_{t_k})$ conditionally to $\{f_s\}_{s\in \Z}$ is equal to $\Pi_{j=1}^k \, f_{t_j}^n$.
\end{itemize}
\end{itemize}
To ensure the existence of a stationary PDF-valued random process ${f_t}_{t\in \mathbb{Z}}$ satisfying Model~\eqref{eq:1}, we require the supplementary hypothesis:
\begin{itemize}
\item[\refstepcounter{hypothesis} \label{hypo:regression} (H\thehypothesis)] It exists $C\in (0, 1)$ such that, for all $f$ and $g$ in $\mathcal{B}_2(I)$, $\| m(f) \ominus m(g) \|_\mathcal{B} \, \leq \, C \, \| f \ominus g \|_\mathcal{B}$.
\end{itemize}
\begin{prop}~\label{prop:dependence}
\begin{itemize}
\item[(i)] Under (H\ref{hypo:regression}), there exists a stationary PDF-valued random process that satisfies the nonparametric model (\ref{eq:1}) and the error process $\{\epsilon_t\}_{t\in\Z}$ is necessarily $\rho$-mixing,
\item[(ii)] $\{\bX_t\}_{t\in \Z}$ is a stationary $\rho$-mixing process.
\end{itemize}
\end{prop}
The proofs of these results are postponed to the Appendix (see~\ref{app:Proposition}). There are several reasons to consider (H\ref{hypo:regression}). Firstly, Proposition~\ref{prop:dependence} shows the importance of this assumption to prove the existence of a stationary PDF-valued random process consistent with the nonparametric regression model. Secondly, the Lipschitz-type condition is typical in the general nonparametric setting, meaning that the regression is sufficiently smooth. With this assumption, it is possible to control the rate of convergence of the bias \citep[see, e.g.,][for the function-on-function nonparametric regression estimator in the standard space of square integrable functions]{FLTV11, FKV12}.
\begin{rem} 
The framework can be extended to accommodate random vectors $\bX_t$'s of random lengths $n_t$'s, where for each $t$, $\bX_t = (X_{t,1}, \ldots, X_{t,n_t})$ and $n_t$ is a random integer. The main asymptotic results given in the next section continue to hold in this more general setting (see Appendix~\ref{app:extension}).
\end{rem}

\subsection{Asymptotic properties}\label{sec:2.2}

This section is devoted to the theoretical properties of our density's functional kernel estimator. We first introduce the notation and additional assumptions that we need to derive the asymptotic behaviour of $\widehat{m}(f)$. 
\begin{itemize}
\item[\refstepcounter{hypothesis} \label{hypo:mixing} (H\thehypothesis)] There exists a constant $a>1$ such that the $\rho$-mixing coefficient satisfies the condition $\rho_f(k) = O(k^{-a})$,
\item[\refstepcounter{hypothesis} \label{hypo:estdensity} (H\thehypothesis)] Let $n = n_N$ be a sequence depending on $N$ such that $n$ goes to infinity with $N$; one assumes the existence of a sequence $\delta_N$ tending to 0 as $N$ goes to infinity such that, for any $t$, $\E \left( \| \widehat{f}_t \ominus f_t \|^2_\mathcal{B} \right) = O(\delta_N^2)$,
\item[\refstepcounter{hypothesis} \label{hypo:sbp} (H\thehypothesis)] For any deterministic $f$ in $\mathcal{B}_2(I)$, the quantity $\pi_f(h_{\text{reg}}) \coloneqq P\left( \| f_1 \ominus f \|_\mathcal{B} < h_{\text{reg}} \right)$ is strictly positive and it exists $C>0$ and $C'>0$, $C \pi_f(h_{\text{reg}}) \leq \pi_f\left(h_{\text{reg}} + o(h_{\text{reg}})\right) \leq C' \pi_f(h_{\text{reg}})$,
\item[\refstepcounter{hypothesis} \label{hypo:asymptotics} (H\thehypothesis)] $h_{\text{reg}}$ tends to 0 with $N$,  $N^{a/(1+a)} \, \pi_f(h_{\text{reg}})$ goes to infinity with $N$, and it exists $b$ with $1/3 < b < 1$ such that  $\delta_N^{2b} = o\left( \pi_f(h_{\text{reg}}) \right)$ and $\delta_N^{1-b} = o\left( h_{\text{reg}} \right)$,
\item[\refstepcounter{hypothesis} \label{hypo:kernel} (H\thehypothesis)] The kernel function $K_{\text{reg}}$ is Lipschitz and positive on its support $[0,\, 1]$.
\end{itemize}
(\prettyref{hypo:mixing}) imposes an arithmetic rate of decay on the mixing coefficients of the PDF-valued random process ${f_t}_{t\in\mathbb{Z}}$. This assumption simplifies the convergence rate, although a more general formulation involving the mixing coefficients $\rho_f(N)$ can also be derived. (\prettyref{hypo:estdensity}) relates the size $n$ of the sample $X_{t,1}, \ldots, X_{t,n}$ to the number $N$ of random PDFs. It requires that the mean integrated squared error of the kernel density estimator $\widehat{f}_t$, $ \text{MISE}(\widehat{f}_t) \coloneqq \E \left( \| \widehat{f}_t \ominus f_t \|_\mathcal{B}^2 \right) $, tends to zero when $n$ (i.e. $N$) goes to infinity. The sequence $\delta_N^2$ can be interpreted as an upper bound of $\text{MISE}(\widehat{f}_t)$ in the Bayes Hilbert space.
Hypothesis (\prettyref{hypo:sbp}) assumes that the behaviour of the small ball probability $\pi_f(h_{\text{reg}})$ is not sensitive to variations in radius negligible with respect to the bandwidth $h_{\text{reg}}$ (see \citealp[][]{LS01} for a survey on the small ball probability). According to the isometry property of the clr transformation (see Section \ref{sec:2.0}), $\pi_f(h_{\text{reg}}) = P( \| X - x \|_2 < h_{\text{reg}})$ with $X = \text{clr}(f_1)$ and $x=\text{clr}(f)$ in $\mathcal{L}^2_0(I)$;  the class of $\mathcal{L}^2_0(I)$-valued random functions satisfying (\prettyref{hypo:sbp}) encompasses Gaussian-type random processes (see \citealp{FZF19}). (\prettyref{hypo:asymptotics}) relates the asymptotic behaviour of $h_{\text{reg}}$, $\pi_f(h_{\text{reg}})$, $N$ and $\delta_N^2$, the upper bound of $\E \left( \| \widehat{f}_t \ominus f_t \|_\mathcal{B}^2 \right)$ for any $t$. As $\pi_f(h_{\text{reg}})$ tends to zero with $N$ faster than $h_{\text{reg}}$, it is reasonable to require that $2b > 1-b>0$, which results in $1/3 < b < 1$. With (\prettyref{hypo:kernel}) and for any $q>0$, $\E \left\{K_{\text{reg}}^{q} ( h^{-1} \, \| f_t \ominus f \|_\mathcal{B} ) \right\}$ is bounded below and above by a quantity of order $\pi_f(h_{\text{reg}})$. For simplicity, we use a basic hypothesis on the kernel function. Nevertheless, it is possible to obtain the more accurate result $\E \left\{K_{\text{reg}}^{q} (h_{\text{reg}}^{-1} \, \| f_t \ominus f \|_\mathcal{B} )\right\} \, = \, C_{f,q} \pi_f(h_{\text{reg}}) \{ 1 + o(h_{\text{reg}})\}$ as soon as the kernel function $K_{\text{reg}}$ is continuously differentiable on $(0, 1)$ with $K_{\text{reg}}(1)>0$, and for all $s \in (0,1)$, $K_{\text{reg}}'(s)\leq 0$ and the ratio $\pi_f(h_{\text{reg}} \, s) / \pi_f(h_{\text{reg}})$ tends to some positive value as $h_{\text{reg}}$ tends to 0.

Once notation and assumptions are introduced, we can provide our main result by giving the pointwise convergence rate of our density-on-density kernel regression estimator. 

\begin{thm} \label{theorem} If (\prettyref{hypo:dependence})-(\prettyref{hypo:kernel}) are fulfilled, then, for any deterministic $f$ in $\mathcal{B}_2(I)$,
$$
\left\| \widehat{m}_N(f) \ominus m(f) \right\|_\mathcal{B} \, = \, O(h_{\text{reg}}) + O\left(\delta_N \, \pi_f(h_{\text{reg}})^{-1/2}\right) + O_P\left( \left\{N^{a/(1+a)} \, \pi_f(h_{\text{reg}})\right\}^{-1/2}\right).
$$
\end{thm}
The first two terms correspond to the bias part. $O(h_{\text{reg}})$, which is not surprising \citep[see the bias part of Theorem 6.11 p. 80 with $\beta=1$ in][]{FV06}, is derived from the regularity assumption acting on the regression operator $m$ (see (\prettyref{hypo:regression})). The second term, which is not usual, involves $\delta_N$ and the small ball probability $\pi_f(h_{\text{reg}})$. This is the price to pay when the random PDFs are not directly observed and are approximated by means of some estimating procedure, where $\delta_N$ is the upper bound of the root $\text{MISE}$ of $\widehat{f}_t$ for any $t$. For the variance part, remember that $a$ controls the asymptotic behaviour of the mixing coefficient of the sequence $\{f_t\}_{t \in \Z}$. So, the impact of the dependence model appears directly in the exponent of the sample size $N$ of the random densities. In the situation of a sample of independent explanatory functional random variables, we would get $N$ instead of $N^{a/(1+a)}$ \citep[see the variance part of Theorem 6.11, p. 80 in][where the $\log n$ vanishes when considering the convergence in probability]{FV06}. 
\begin{rem}
“The proof of Theorem \ref{theorem} (see Section \ref{app:B} in Appendix B) does not use the fact that the Bayes Hilbert space is isometric to $\mathcal{L}_0^2(I)$. Consequently, as soon as the unobserved objects can be estimated consistently, Theorem \ref{theorem} remains valid for non-Euclidean spaces (i.e., spaces not isometric to any Euclidean space).
\end{rem}

\noindent Let us now focus on the kernel density estimator 
\begin{equation}
\widehat{f}_t(x) = (n \, h_t)^{-1} \sum_{i=1}^{n} K_{\text{kde}}\left\{h_t^{-1} ( x - X_{t,i} ) \right\}, \label{eq:density_est}
\end{equation}
with $h_t$ a bandwidth tending to zero when $n$ goes to infinity and $K_{\text{kde}}$ is a standard univariate kernel function with support of $K_{\text{kde}}$ equals to $[-1,1]$ (i.e. symmetric, integrates to 1, and $\int_{-1}^1 u^2 \, K_{\text{kde}}(u) \, du < \infty$). When $f_t$ is deterministic, the asymptotic approximation of $\text{MISE}(\widehat{f}_t)$ based on the iid sample $X_{t,1}, \ldots, X_{t,n}$ is well known: $\text{MISE}(\widehat{f}_t)  \, = \, O\left(h_t^4 \right) + O\left(n^{-1}h_t^{-1} \right)$ \citep[for more details on MISE, see][among others]{Rosenblatt56, Epanechnikov69, Wegman72, Nadaraya74, MW92, WJ94, Hansen05}. This result requires that the second derivative $f_t''$ of the true density function $f_t$ is continuous and satisfies $\int ({f_t"})^2 < \infty$. The extension of this result to a random density $f_t$ with the probabilistic environment defined in (H\ref{hypo:dependence}) is obtained thanks to Theorem~\ref{theo2:rate} (see~\eqref{eq:asymp_kde9} in Appendix~\ref{app:proof_theo2}); it requires the almost sure (a.s.) version of these regularity assumptions: 

\begin{itemize}
\item[\refstepcounter{hypothesis} \label{hypo:densityderiv} (H\thehypothesis)] $P\left(\left\{ \omega\in \Omega, \, f_1''(\omega,.) \text{ is continuous and } \int_I f_1''(\omega,.)^2  < \infty \right\}\right)=1$.
\end{itemize}
The following result simplifies the rate of convergence given in {\sc Theorem}~\ref{theorem} when the random densities are estimated with the kernel density estimator \eqref{eq:density_est}. 
\begin{thm}\label{theo2:rate}
If, in addition of (\prettyref{hypo:dependence})-(\prettyref{hypo:densityderiv}), one has
\begin{itemize}
\item[\refstepcounter{hypothesis} \label{hypo:densityasym} (H\thehypothesis)] For any $t$, it exists $q > 5 a / \{ 4 ( 1 + a ) \}$ such that $n \asymp N^q$ and $h_t \asymp N^{-q/5}$, 
\end{itemize}
\begin{itemize}
\item[(i)]  $\displaystyle \E\| \widehat{f}_t - f_t \|_\mathcal{B}^2 = O\left(N^{-4q/5} \right)$,
\item[(ii)] $\displaystyle \left\| \widehat{m}_N(f) \ominus m(f) \right\|_\mathcal{B} \, = \, O(h_{\text{reg}}) + O_P\left( \left\{N^{a/(1+a)} \, \pi_f(h_{\text{reg}})\right\}^{-1/2}\right)$,
\end{itemize}
\end{thm}
where $u \asymp v$ means that $u$ and $v$ are equivalent in the sense $u = O(v)$ and $v = O(u)$. (\prettyref{hypo:densityasym}) corresponds to a particular case where $\delta_N^2$, the upper bound of $\E \left( \| \widehat{f}_t \ominus f_t \|_\mathcal{B}^2 \right)$, is negligible with respect to $N^{-a/(1+a)}$ so that the term $O\left(\delta_N \, \pi_f(h_{\text{reg}})^{-1/2}\right)$ vanishes. The proof of Theorem \ref{theo2:rate} is postponed to the Appendix (see Section~\ref{app:proof_theo2}).
\section{Finite-sample properties}

\subsection{Implementation}\label{sec:3.1}

Since actual densities in practical applications are not often observable, we work with estimated densities $\widehat{f}_1, \ldots, \widehat{f}_{N+1}$ which are outputs of the kernel density estimator defined with (\ref{eq:density_est}) where $K_{\text{kde}}(\cdot)$ stands for a kernel function and $h_t$ denotes a bandwidth for period $t$. We consider the truncated Gaussian kernel.

In our nonparametric density-on-density regression, we consider how to select the optimal bandwidths for estimating the nonparametric regression operator in~\eqref{eq:1} and for estimating densities in~\eqref{eq:density_est}. To estimate densities from observed data, we use a kernel density estimator with the bandwidth selected by Silverman's rule-of-thumb (ROT) \citep[see also][]{KMP+19}. \cite{KMP+19} used Silverman's ROT to select the bandwidth, which leads to $\widehat{h}_t = 2.34\times \widehat{\sigma}_t \times n^{-1/5}$,
where $\widehat{\sigma_t}$ denotes the sample standard deviation of $X_{t,i}$, $i=1,\dots,n$. 

To estimate the bandwidth for the nonparametric regression operator, we can use generalised cross-validation (GCV) to select the bandwidth automatically \citep[see, e.g.,][]{FVV12}. The GCV aims to minimise an overall squared loss function between the estimated and observed responses among the data in the training sample.




\subsection{Monte-Carlo simulation studies}\label{sec:3.3}


We conduct a series of simulation studies to examine the finite-sample performance of the nonparametric density-on-density regression described in Section~\ref{sec:2}. Our data generating process consists of the following steps:
\begin{enumerate}
\item[1)] When generating the model \(f_{t+1} = m(f_t)\oplus \epsilon_t\), a first step is to be able to simulate \(\epsilon_0, \epsilon_1,\ldots\), the probability density functions playing the role of model errors. By using the clr transformation, the density-on-density regression model is equivalent to \(\mbox{clr}(f_{t+1}) = \mbox{clr} \circ m(f_t) + \mbox{clr}(\epsilon_t)\). Instead of generating directly the errors \(\epsilon_t\) in the Bayes Hilbert space \(\mathcal{B}_2(I)\), it could be more convenient to simulate functional errors \(\eta_t = \mbox{clr}(\epsilon_t)\) in the corresponding Hilbert space~\(\mathcal{L}^2_0(I)\).
\item[2)] According to the properties of the clr transformation and the zero-mean constraint in \(\mathcal{B}_2(I)\) of \(\epsilon_t\) given \(f_t\), the model error \(\eta_t\) has to fulfill two constraints: \(\E(\eta_t|f_t) = 0\), and for all \(u \in \)  \(I\), \(\displaystyle\int_I \eta_t(u) \, du = 0\). A way to generate such \(\eta_t\)'s is to use the trigonometric functions \(\phi_1, \phi_2, \ldots\) defined on \(I=[-1,\, 1]\): \(\forall k=1,2\ldots, \ \phi_{2k-1}(u) = \cos(k \pi \, u)\) and \(\phi_{2k}(u) = \sin(k \pi \, u)\). One can remark that the \(\phi_k\)'s integrate to 0 for any non-null integer \(k\). For any \(u \in [-1, \, 1]\), and any subset \(\mathcal{K}\) of nonnull integers, set \[ \eta_t(u) \, = \, \sum_{k\in \mathcal{K}} A_{tk} \, \phi_k(u). \] If \(\E(A_{tk}|f_t)=0\), then \(\E(\eta_t|f_t)=0\) and the integral property of the Fourier basis elements results in \(\int_{-1}^1\eta_t(u) \, du = 0\). For instance, one can set \[ \eta_t(u) \, = \, A_{t1} \cos(\pi \, u) + A_{t2} \sin(\pi \, u) + A_{t3} \cos(2 \pi u) + A_{t4} \sin(2 \pi \, u) + A_{t5} \cos(3 \pi \, u), \] where \(A_{tj}\)'s are iid zero-mean rrv. In Figure~\ref{fig:sim_model_error}, we present such model errors \(\eta_t\) with \(A_{t1}\)'s,\dots,\(A_{t5}\)'s \(\overset{iid}{\sim}\) \(N(0, \sigma^2)\).
\begin{figure}[!htbp]
\centering
{\includegraphics[width=8.2cm]{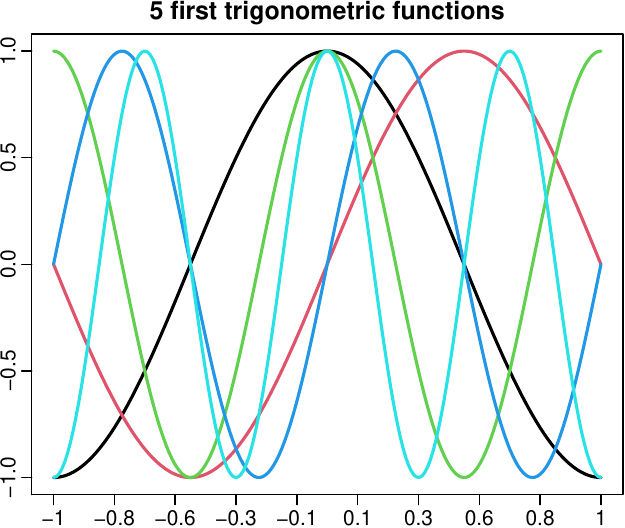}}
\quad
{\includegraphics[width=8.2cm]{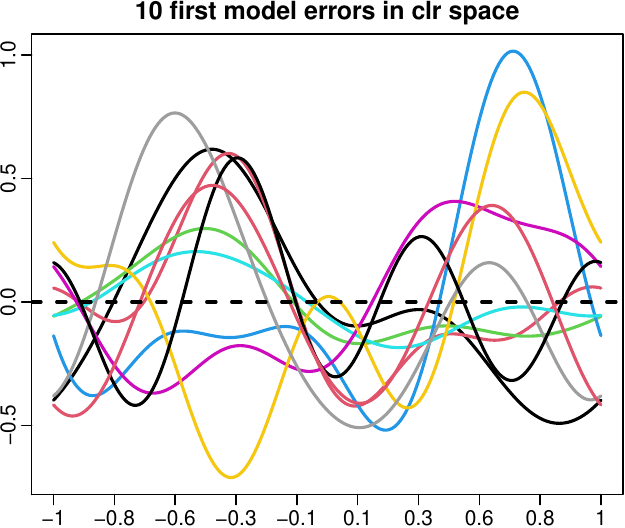}}
\caption{\small{The basis functions we considered are the first five trigonometric functions. By multiplying them with random coefficients simulated from a Gaussian distribution, we obtain the model error in the clr space, $\eta_t(u)$.}}\label{fig:sim_model_error}
\end{figure}

Transform back to \(\epsilon_t\)'s with the inverse of the clr transformation: \(\epsilon_t = \mbox{clr}^{-1}(\eta_t)\). In Figure~\ref{fig:sim_model_err_PDF}, we present 10 first model errors in the Bayes space.
\begin{figure}[!htbp]
\centering
\includegraphics[width=9cm]{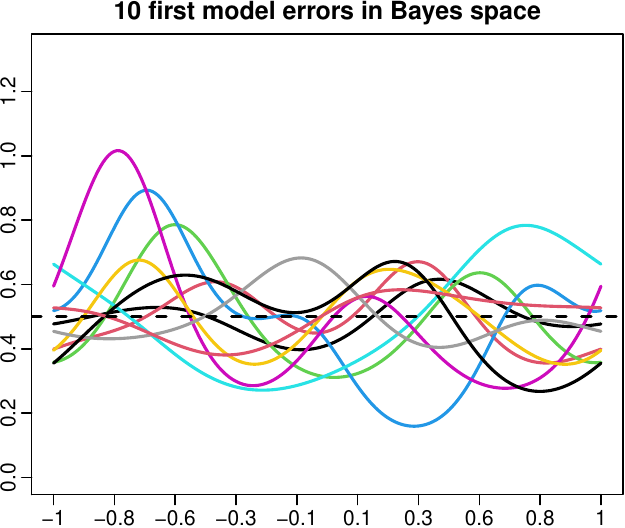}
\caption{\small By taking the inverse clr transformation of the model error in Figure~\ref{fig:sim_model_error}, we obtain the model error in Bayes space.}\label{fig:sim_model_err_PDF}
\end{figure}

\item[3)] Let us define the regression operator \(m: \, \mathcal{B}_2(I) \rightarrow \mathcal{B}_2(I)\) such that \[\displaystyle m(f)(y) \, = \, (1-\rho_0)^{-1} \left\{ \int_{\underline{b}}^{\overline{b}} f(x) \, g\left(\frac{y -\rho_0 \, x}{1 - \rho_0}\right) \, dx \right\} 1_{[-1,\, 1]}(y)\] with \(f\) and \(g\) two given probability densities in \(\mathcal{B}_2(I)\), \(0<\rho_0<1\) and where $\underline{b} = \max[-1$, $\rho_0^{-1}(y + \rho_0 - 1)]$ and $\overline{b} = \min[1, \rho_0^{-1}(y - \rho_0 + 1)]$. Depending on the value of $\rho_0$, the function support may vary for the regression mean function. This regression operator has a nice interpretation in terms of convolution. Suppose that for any \(0<\rho_0<1\), \(Z = \rho_0 X + (1-\rho_0) Y\) with \(X\) and \(Y\) two independent rrv and let \(f_Z\) (resp. \(f_X\)) be the PDF of \(Z\) (resp. \(X\)). Then, we have \(f_Z \, = \, m(f_X)\) and $\text{Corr}(Z, \, X) = \rho_0$. The regression operator \(m(.)\) corresponds to the transformation resulting from a convex combination of \(X\) with another rrv \(Y\). Based on this remark, we propose the following scheme for simulating our time series of PDFs in the Bayes Hilbert space: 
\begin{equation*} 
f_{t+1} \, = m(f_t) \oplus \epsilon_t,
\end{equation*}
where the \(\epsilon_t\)'s are the model errors in the Bayes space (built previously):
\begin{inparaenum}
\item[1)] \(m(f_t)\) represents the PDF of $X_{t+1} = \rho_0 X_t + (1-\rho_0) Y_{t+1}$ when $f_t$ is the PDF of $X_t$,
\item[2)] \(f_{t+1}\) is the obtained PDF when adding some model error \(\epsilon_t\) to \(m(f_t)\).
\end{inparaenum}
\item[4)] Set $Y_t \, \sim \, g_t = TN(\mu_t, \nu^2)$ where \(TN(\mu_t, \nu^2)\) is the truncated normal distribution over~$[-1,1]$ derived from the normal distribution \(N(\mu_t, \nu^2)\) where, for some constant \(T>0\), \(\mu_t = \cos(2\pi t/T)\). 
\item[5)] Set \(f_1 \equiv g_1\) where \(g_1 = TN(\mu_1, \nu^2)\) with \(\mu_1 = \cos(2\pi /T)\). We compute  \(f_2 = m(f_1) \oplus \epsilon_1\). Then, we iterate to build the whole time series of densities \(f_1,\, f_2 = m(f_1)\oplus\epsilon_1,\ldots,\, f_N=m(f_{N-1})\oplus\epsilon_{N-1}\).
\end{enumerate}

\begin{figure}[!htbp]
\centering
\includegraphics[width=8.7cm]{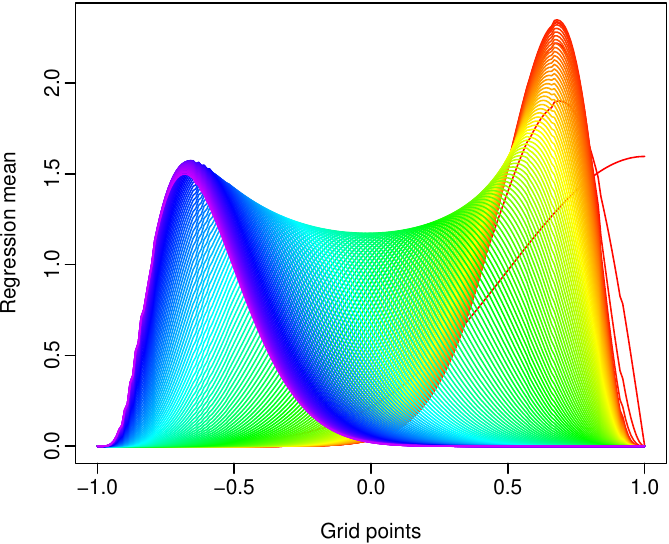}
\quad
\includegraphics[width=8.7cm]{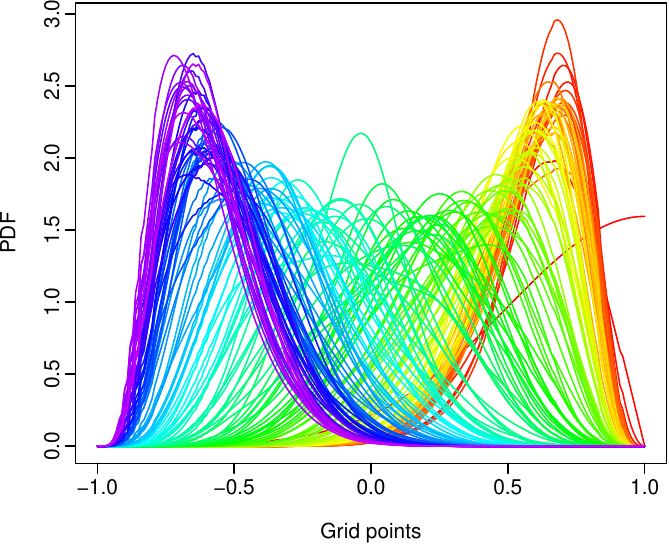}
\caption{\small 150 simulated densities with 201 grid points bounded between -1 and 1. The four tuning parameters are chosen as: $\sigma=0.1$, $\rho=0.5$, $\nu=0.5$ and $T=150$.}\label{fig:4_png}
\end{figure}

In Figure~\ref{fig:4_png}, we present a perspective plot of a simulated example with $150$ curves. We consider 201 grid points within a function support range $[-1,1]$. There are four tuning parameters to choose, namely $\sigma$ in step 2), $\rho_0$ in step 3), $\nu$ and $T$ in step 4). Let $\sigma=0.1$, $\rho_0 = 0.5$, $\nu=0.5$ and $T=150$.

We divide the simulated data into a training sample and a testing sample comprised of the last 50 densities. To evaluate the finite-sample performance, we compute the Kullback-Leibler divergence (KLD) \citep{KL51}. The KLD is intended to measure the loss of information when we choose an approximation. For two density functions, the discrete version of the Kullback-Leibler divergence is defined as
\begin{align*}
\text{KLD}_m &= \frac{1}{50}\sum^{50}_{i=1}\left\{D_{\text{KL}}[f_i^{m}(s)||\widehat{f}_i^{m}(s)] + D_{\text{KL}}[\widehat{f}_i^{m}(s)||f_i^{m}(s)]\right\} \\
\overline{\text{KLD}} &= \frac{1}{100}\sum^{100}_{m=1} \text{KLD}_m,
\end{align*}
where $f_i^{m}(s)$ is the $i$\textsuperscript{th} observation in the testing sample, and $\widehat{f}_i^m(s)$ is the one-step-ahead forecast of $f_i^{m}(s)$ for the $m$\textsuperscript{th} replication. 

\begin{figure}[!htbp]
\centering
\includegraphics[width = 5.6cm]{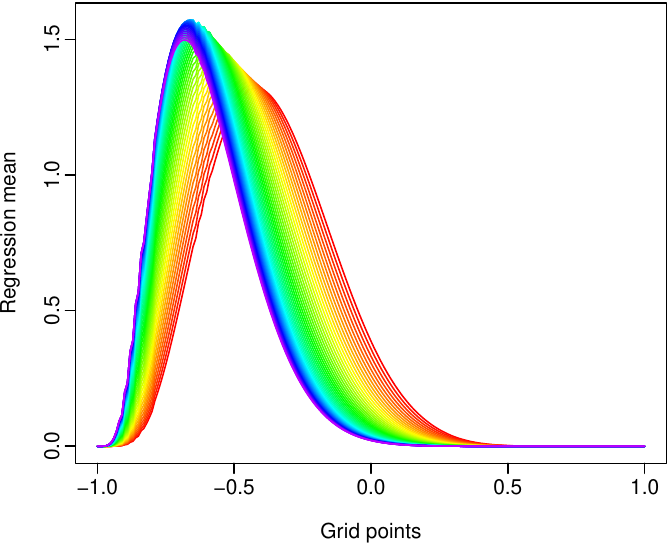}
\quad
\includegraphics[width = 5.6cm]{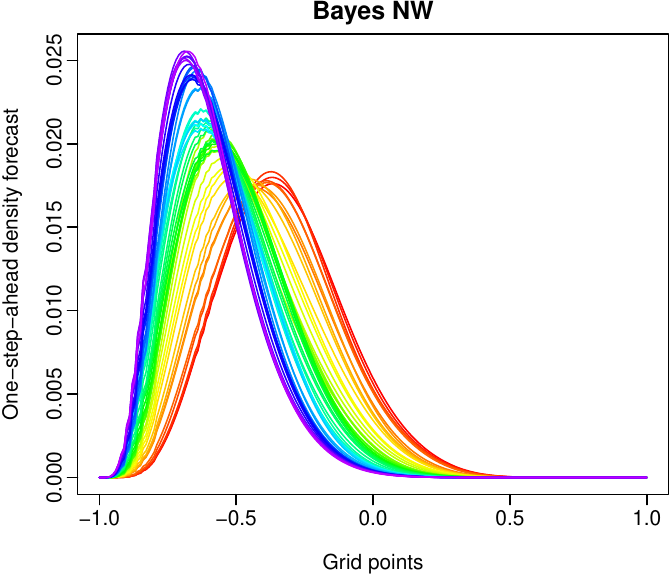}
\quad
\includegraphics[width = 5.6cm]{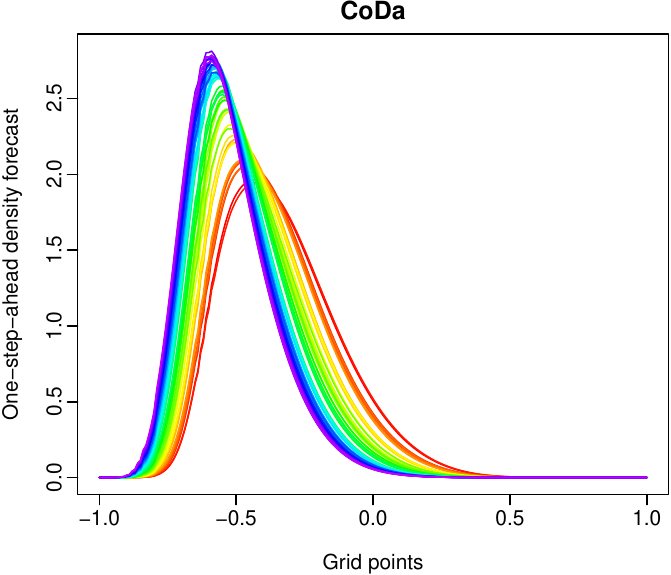}
\\
\includegraphics[width = 5.6cm]{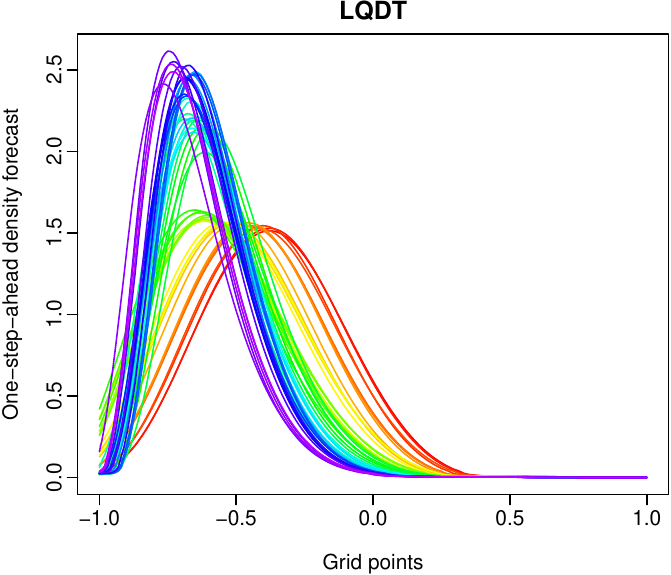}
\quad
\includegraphics[width = 5.6cm]{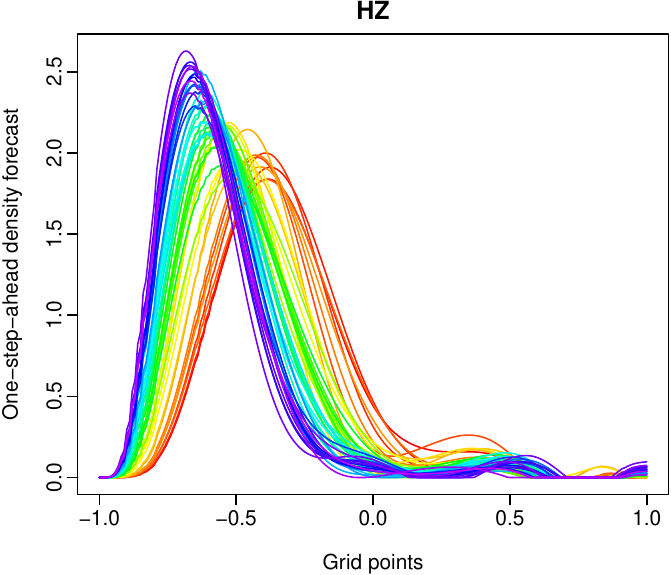}
\quad
\includegraphics[width = 5.6cm]{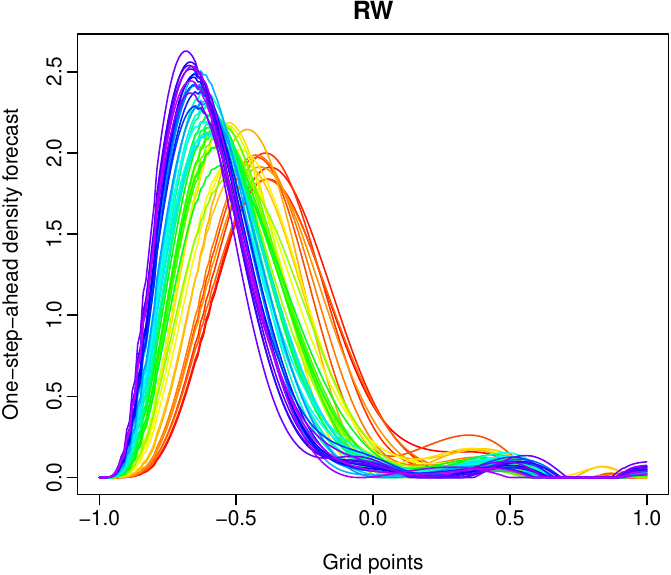}
\caption{\small One simulated example with a sample size of $T=150$ and 201 grid points between -1 and 1. While the first 100 densities are used as the initial training sample, we produce iterative one-step-ahead forecast densities via an expanding window scheme. HZ stands for \citeauthor{HZ18}'s \citeyearpar{HZ18} method.}\label{fig:2_demo}
\end{figure}

For one out of 100 replications, we generate a sample of densities under different tuning parameters. With the training samples, we produce one-step-ahead forecast densities using the nonparametric density-on-density regression with the Bayes NW estimator, the CoDa method, the LQDT, functional principal component regression of \cite{HZ18}, and random walk. In Figure~\ref{fig:2_demo}, we present one replication of holdout densities and their forecasts obtained by the five methods. For this example, we observe that the nonparametric density-on-density regression produces the best results, followed closely by the random walk. The functional principal component regression of \cite{HZ18} suffers from the well-known tail problem when density forecasts can even be negative.

To further assess the overall forecast accuracy, we repeat the simulation data generating process 100 times, each with different pseudo-random seeds. In Table~\ref{tab:sim}, we present the one-step-ahead averaged KLDs between the actual holdout samples and their corresponding forecasts obtained from the five methods. In addition, we present results using an expanding window approach in which the size of the training samples increases. As measured by the $\overline{\text{KLD}}$ criterion, the forecast accuracy relies heavily on the noise-to-signal ratio, characterised by the dependence parameter $\rho_0$ and the standard deviation $\sigma$ in the error term of the AR(1) structure. As the dependence parameter $\rho_0$ decreases from 0.75 to 0.25, we observe a decrease of $\overline{\text{KLD}}$ based on 100 replications. As the standard deviation in the error term of the AR(1) structure increases from 0.1 to 1, we also observe an increase of $\overline{\text{KLD}}$. As the sample size increases from $150$ to 550, we observe a general increase of $\overline{\text{nsr}}$ which does not translate to an increase of $\overline{\text{KLD}}$.
\begin{table}[!htb]
\centering
\tabcolsep 0.2in
\caption{\small{One-step-ahead forecast accuracy of the nonparametric density-on-density regression, simulated data with 100 replications. We consider three choices of $(\sigma, \rho_0)$ with sample sizes of $T=150, 350$ and $550$. The last 50 densities are the testing sample, while the remaining densities are the training sample. For the 100 replications, we compute the average noise-to-signal ratio, denoted by $\overline{\text{nsr}}$. HZ stands for \citeauthor{HZ18}'s \citeyearpar{HZ18} method.} \label{tab:sim}}
\begin{tabular}{@{}llrrrrrr@{}}
\toprule
&  & & \multicolumn{5}{c}{Method} \\
\cmidrule{4-8}
$(\sigma, \rho_0)$ & $n$ & $\overline{\text{nsr}}$ & Bayes NW & CoDa & LQDT & HZ & RW \\
\midrule
$(0.10, 0.75)$   & 150 & 0.0398 & 0.0137 & 1.2085 & 3.7018 & 0.6377 & \textBF{0.0115} \\
			& 350 & 0.0918 & \textBF{0.0079} & 0.4419 & 3.5817 & 0.0818 & 0.0108 	\\
			& 550 & 0.1439 & 0.0209 & 0.1213 & 0.0173 & 0.1772 & \textBF{0.0197} 	\\
\\
$(0.10, 0.50)$   & 150 & 0.0219 &  \textBF{0.0069} & 0.1339 & 0.0880 & 0.0963 & 0.0097	\\
			& 350 & 0.0353 & \textBF{0.0035} & 0.0593 & 0.0155 & 0.0389 & 0.0094 	\\
			& 550 & 0.0514 & \textBF{0.0082} & 0.0108 & 0.0119 & 0.0754 & 0.0125 	\\
\\
$(0.10, 0.25)$   & 150 & 0.0272 & \textBF{0.0069} & 0.0170 & 0.2433 & 0.0628 & 0.0098	\\
			& 350 & 0.0340 & \textBF{0.0016} & 0.0079 & 0.1584 & 0.0516 & 0.0098		\\
			& 550 & 0.0434 & 0.0046 & \textBF{0.0032} & 0.0880 & 0.0501 & 0.0122 	\\
	\\	
$(0.50, 0.75)$   & 150 & 0.9703 & \textBF{0.1638} & 1.1957 & 3.1611 & 0.6668 & 0.2371 \\
			& 350 & 2.2291	& \textBF{0.1200} & 0.4611 & 3.3656 & 0.3557 & 0.2334 \\	
			& 550 & 3.4945	& \textBF{0.1376} & 0.1636 & 0.1967 & 0.4982 & 0.2717 \\
\\
$(0.50, 0.50)$   & 150 & 0.5338 & \textBF{0.0784} & 0.1924 & 0.1072 & 0.2682 & 0.2124 \\
			& 350 & 0.8582	& \textBF{0.0434} & 0.0901 & 0.1528 & 0.2076 & 0.2134 \\
			& 550 & 1.2480	& 0.0540 & \textBF{0.0396} & 0.1332 & 0.2796 & 0.2423 \\
\\
$(0.50, 0.25)$   & 150 & 0.6640	& 0.0522 & \textBF{0.0382} & 0.0994 & 0.2334 & 0.2191 \\
			& 350 & 0.8259	& 0.0233 & \textBF{0.0207} & 0.1421 & 0.1564 & 0.2189 \\
			& 550 & 1.0549	& 0.0304 & \textBF{0.0184} & 0.1772 & 0.1880 & 0.2499 \\
\\								
$(1, 0.75)$ & 150 & 3.1793 & \textBF{0.5017}	& 0.6227	& 3.3689	& 0.9971	& 0.7988 \\
		  & 350 & 7.3052 & 0.3878	& \textBF{0.3745}	& 3.3869	& 0.7850	& 0.8163 \\
		  & 550 & 11.4313 & 0.3754 	& \textBF{0.3416}	& 0.5763	& 0.9343	& 0.8222 \\
\\
$(1, 0.50)$ & 150 & 1.7491 & \textBF{0.2599}	& 0.3605	& 0.7276	& 0.4833	& 0.7203 \\
		  & 350 & 2.8124 &  \textBF{0.1561}	& 0.1923	& 0.8041	& 0.4716	& 0.7303 \\
		  & 550 & 4.0825 & 0.1743	& \textBF{0.1165}	& 0.4573	& 0.5774	& 0.7667 \\
\\
$(1, 0.25)$ & 150 & 2.1756 & 0.1720	& \textBF{0.1144}	& 0.7714	& 0.3197	& 0.7273 \\
		  & 350 & 2.7067 & 0.0774	& \textBF{0.0615}	& 0.8935	& 0.3034	& 0.7255 \\
		  & 550 & 3.4510 & 0.1134	& \textBF{0.0572}	& 0.5735	& 0.3942	& 0.7929 \\		  		  	
\bottomrule
\end{tabular}
\end{table}

We observe that the \citeauthor{HZ18}'s \citeyearpar{HZ18} method produces inferior accuracy because it ignores the density constraints. By obeying the density constraints, the remaining four methods improve the forecast accuracy. Among the four methods, the LQDT does not work well when functions have different finite supports, as is evident from the large $\overline{\text{KLD}}$ values. When the value of $\sigma$ is lower, the nonparametric density-on-density regression using the functional Bayes NW estimator is the method chosen with the smallest $\overline{\text{KLD}}$. When the value of $\sigma$ is higher and the value of $\rho_0$ is lower, the CoDa method is preferred.

\subsection{Analysis of a French COVID19 hospitalisation dataset}\label{sec:3.6}

\begin{figure}[!htb]
\centering
\includegraphics[width=16.5cm]{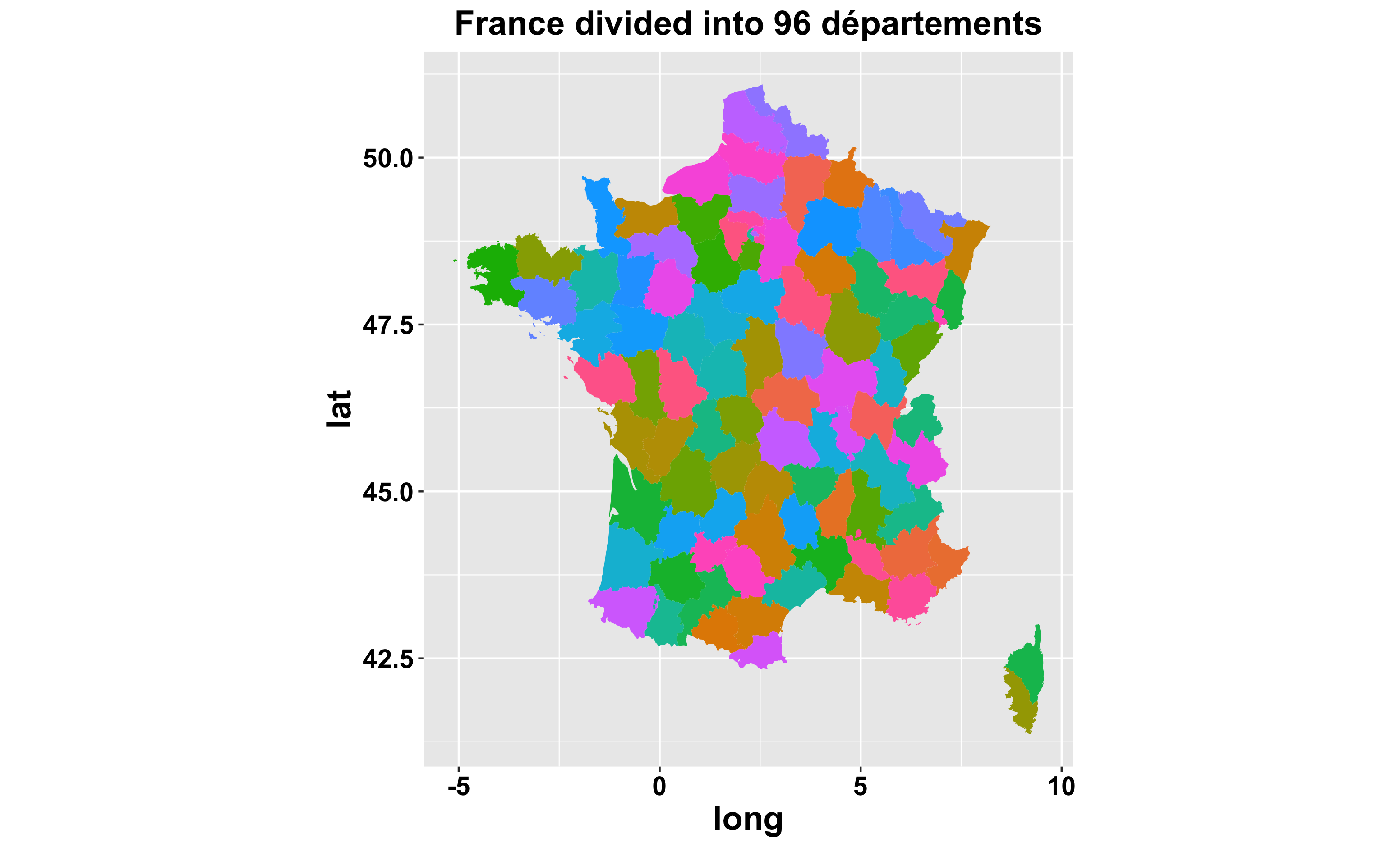}
\caption{A map of French departments}\label{fig:map_fr}
\end{figure}

The French open data portal \url{https://www.data.gouv.fr/fr/} allows the general public to access many data sources. Simple research on this site with the keyword COVID19 leads us to several data sets. In this subsection, we are interested in the dynamics of the pandemic in France at the departmental level and, more particularly, in the hospitalisation data. The French territory is divided into 101 administrative areas named d\'{e}partements. In Figure~\ref{fig:map_fr}, we plot 96 departments located within Europe. 

The French COVID19 data, updated daily, are contained in a .csv file whose name starts with covid-hospit-2023 and can be downloaded \url{https://www.data.gouv.fr/fr/datasets/donnees-hospitalieres-relatives-a-lepidemie-de-covid-19/}. We are interested in studying the number of hospitalisations per 100,000 inhabitants for each department from March 18, 2020, to August 24, 2022. Using cross-sectional data, we collect 96 daily hospitalisation ratios $(\X_{t,1},\X_{t,2},\dots,\X_{t,96})$ for each observed date $t=1,2,\dots,890$. By taking a simple average over days, we can identify departments with minimum and maximum hospitalisation ratios. The department Vend\'{e}e has the minimum hospitalisation ratio, with an average of 8.58 per 100,000 inhabitants. The Territoire de Belfort department has the highest hospitalisation ratio, with an average of 50.52 per 100,000 inhabitants.

\begin{figure}[!htbp]
\centering
\includegraphics[width=10cm]{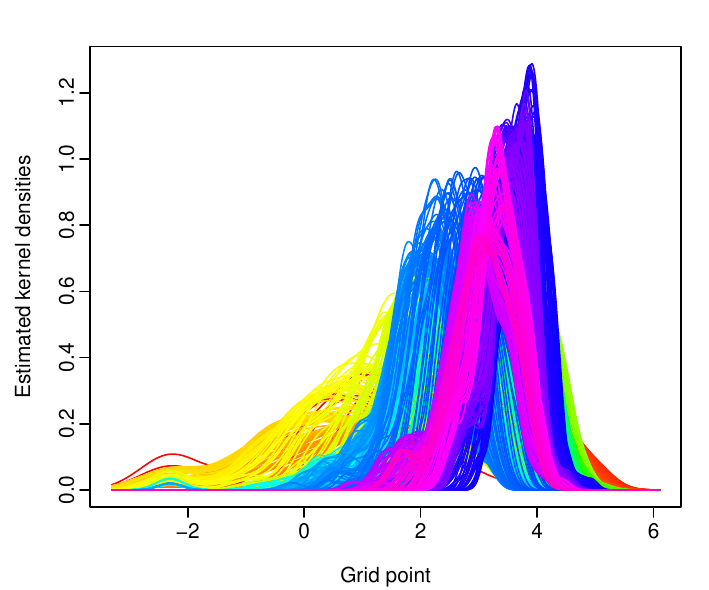}
\caption{A time series of estimated probability density functions via a kernel density estimator with truncated Gaussian kernel and Silverman's ROT for the bandwidth selection. The estimated densities reflect daily COVID19 hospitalisation counts across the 96 departments.}\label{fig:france_densities}
\end{figure}

Since the number of hospitalisation is nonnegative, the nonnegativity constraint can be removed by applying the natural logarithm transformation, denoted by $\ln(\cdot)$. Let $X_{t,i}=\ln(\X_{t,i}+c)$ with $c>0$ for $i=1, 2,\dots,96$, where $c$ is an arbitrary small value such as $c=0.1$. Then, we apply a kernel density estimator to estimate the probability density function $f_t$ based on the sample $\{X_{t,1}, \ldots, X_{t, 96}\}$ without a positive constraint. Computationally, the estimation is implemented via the density function in \Rlogo \ \citep{Team22} using Silverman's ROT bandwidth selection and the truncated Gaussian kernel function. In Figure~\ref{fig:france_densities}, we present the 890 estimated densities.

From the estimated $(\widehat{f}_1, \widehat{f}_2,\dots,\widehat{f}_{890})$, we split the samples into an initial training sample and a testing sample. The initial training sample consists of the first 594 estimated densities, while the testing sample consists of the remaining 296 estimated densities. Using the first 594 estimated densities from March 18, 2020, to November 1, 2021, we produce one-step-ahead forecasts and evaluate its KLD with its holdout testing sample on November 2, 2021. Through an expanding window approach, we increase the training sample by one and use the first 595 estimated densities to produce one-step-ahead forecasts and evaluate its KLD on November 3, 2021. We proceed in the same manner to obtain 296 KLDs corresponding to the testing period.

\begin{figure}[!htb]
\centering
\includegraphics[width=18cm]{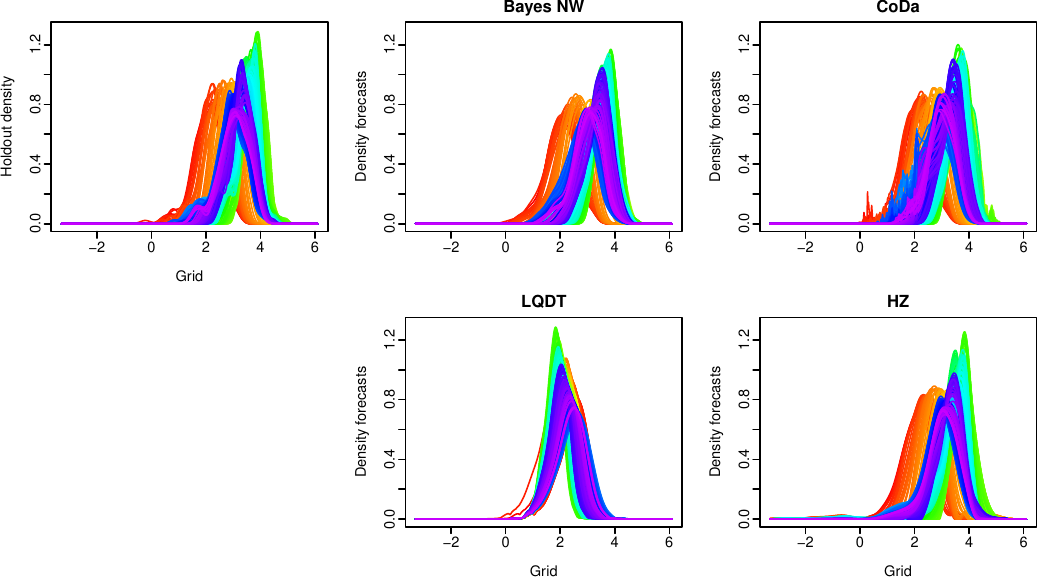}
\caption{\small For the 296 holdout densities, we present the one-step-ahead density forecasts of the four methods based on an expanding window.}\label{fig:french_forecasts}
\end{figure}

In Figure~\ref{fig:french_forecasts}, we present the 296 one-step-ahead density forecasts of the four methods considered. Since the regression mean function (i.e., signal) is unknown, the random walk is not considered a benchmark method. Compared with the holdout densities, we found that nonparametric density-on-density regression with the Bayes NW estimator performs the best. Similar to the CoDa method, it also uses the log-ratio transformation to remove constraints. A univariate functional time series forecasting method or a nonparametric function-on-function is then applied to obtain the forecasts. The forecasts are then back-transformed via the inverse log-ratio transformation. The nonparametric density-on-density regression with the Bayes NW estimator generally produces smoother density forecasts than the ones obtained from the CoDa method. The intuition is that the former is a weighted average of the past densities and tends to be smooth.

Among the methods, the LQDT exhibits a different pattern. Using other time-series forecasting methods, they do not improve the forecast accuracy measured by the KLD. We suspect that the issue may lie in the chosen reference point. Without an adequate reference point, the LQDT would only be defined up to an arbitrary constant.

In Table~\ref{tab:french_covid_table}, we present summary statistics of the KLDs obtained from the four methods. Based on the median and mean values, the nonparametric density-on-density regression with the Bayes NW estimator performs the best among the four methods. 
\begin{table}[!htbp]
\centering
\caption{\small From the 296 KLDs, we present summary statistics for the four methods, as well as the random-walk (RW) method. HZ stands for \citeauthor{HZ18}'s \citeyearpar{HZ18} method.}\label{tab:french_covid_table}
\tabcolsep 0.35in
\begin{tabular}{@{}lrrrrr@{}}
\toprule
 Statistic & Bayes NW & CoDa & LQDT & HZ & RW \\ 
\midrule
  Min. & 0.0041 & 0.0051 & 0.0171 & 0.0176 & 0.0008 \\ 
  1st Qu. & 0.0188 & 0.0197 & 1.0127 & 0.0727 & 0.0185 \\ 
  Median & 0.0293 & 0.0324 & 4.1861 & 0.1747 & 0.0331 \\ 
  Mean & 0.0420 & 0.0476 & 8.4794 & 0.2014 & 0.0450 \\ 
  3rd Qu. & 0.0524 & 0.0554 & 12.8809 & 0.2910 & 0.0561 \\ 
  Max. & 0.5241 & 0.4243 & 33.2268 & 0.9041 & 0.4461 \\ 
\bottomrule
\end{tabular}
\end{table}

\vspace{-.3in}

\subsection{Analysis of age-specific life-table death counts in the United States}\label{sec:3.5}

While it requires a way of estimating densities using a kernel estimator in Section~\ref{sec:3.6}, sometimes, we observe a time series of densities, such as period life table in demography and actuarial studies. A period life table is a table that shows, for each age, the probability that a person of that age will die before the next birthday. In the first age group, the initial number of alive is 100,000, while the remaining number of alive is 0 in the last age group. The sum of life-table death count is 100,000 every year, and the values are nonnegative, so these observations can be viewed as densities normalised to 100,000 rather than 1. By modelling the life-table death counts, we could understand a redistribution of survival probabilities, where deaths at younger ages gradually shift towards older ages. In Figure~\ref{fig:1}, we present rainbow plots of the female and male age distributions of life-table death counts in the U.S. from 1933 to 2023 in a single-year group.

\begin{figure}[!htb]
\centering
\includegraphics[width = 5.6cm]{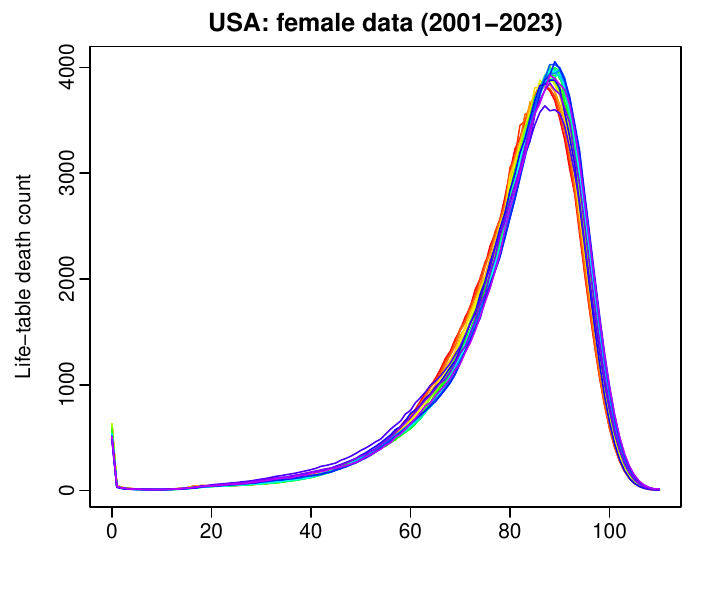}
\quad
\includegraphics[width = 5.6cm]{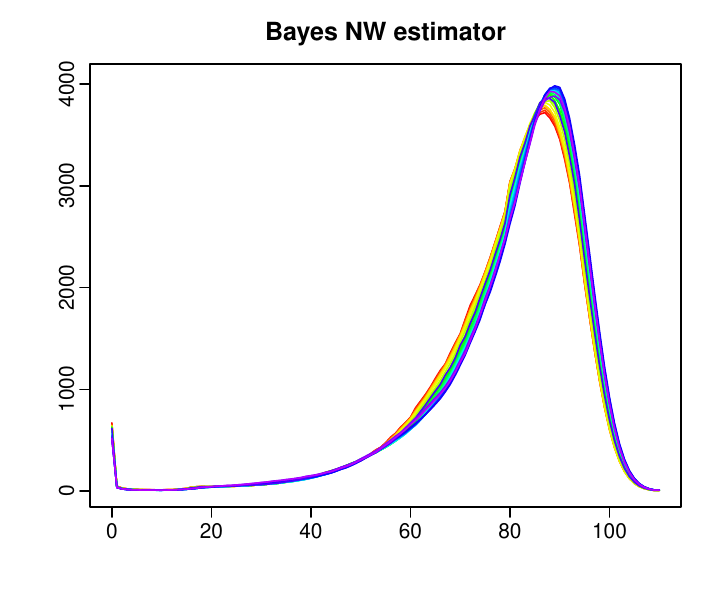}
\quad
\includegraphics[width = 5.6cm]{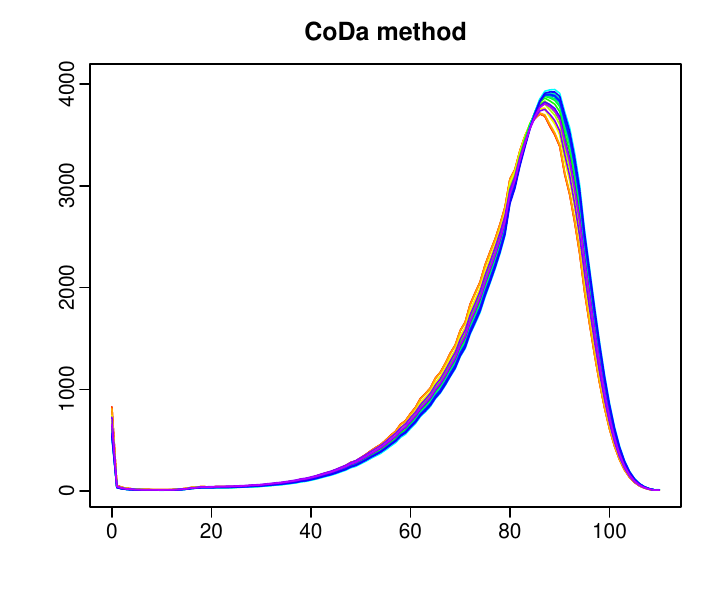}
\\
\includegraphics[width = 5.6cm]{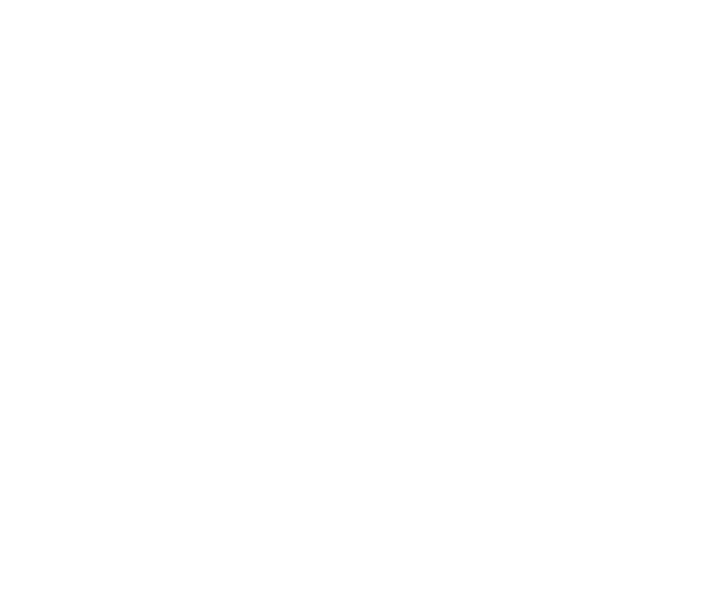}
\quad
\includegraphics[width = 5.6cm]{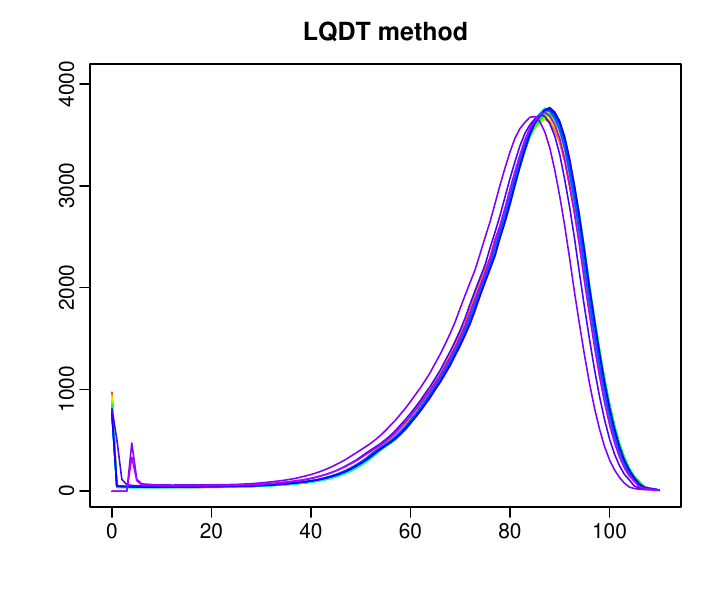}
\quad
\includegraphics[width = 5.6cm]{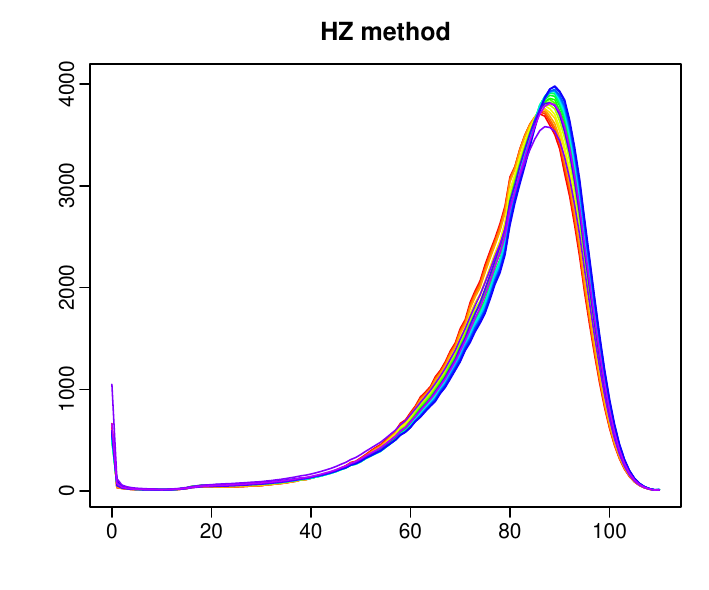}
\\
\includegraphics[width = 5.6cm]{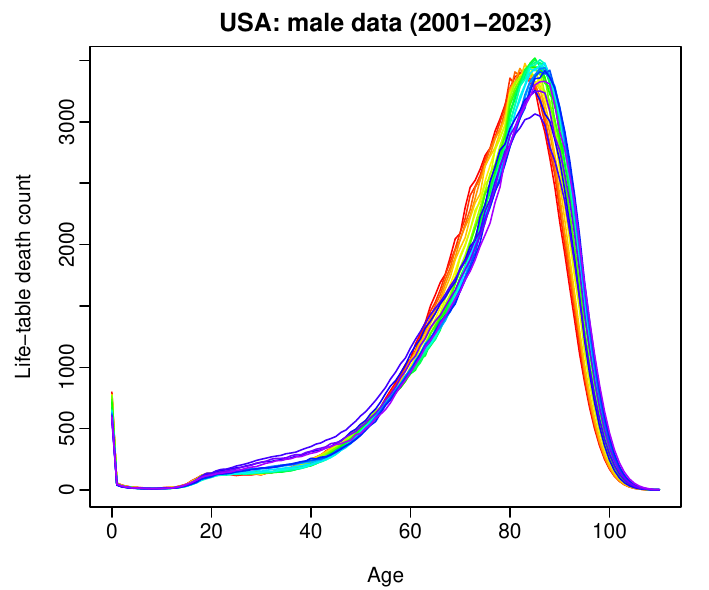}
\quad
\includegraphics[width = 5.6cm]{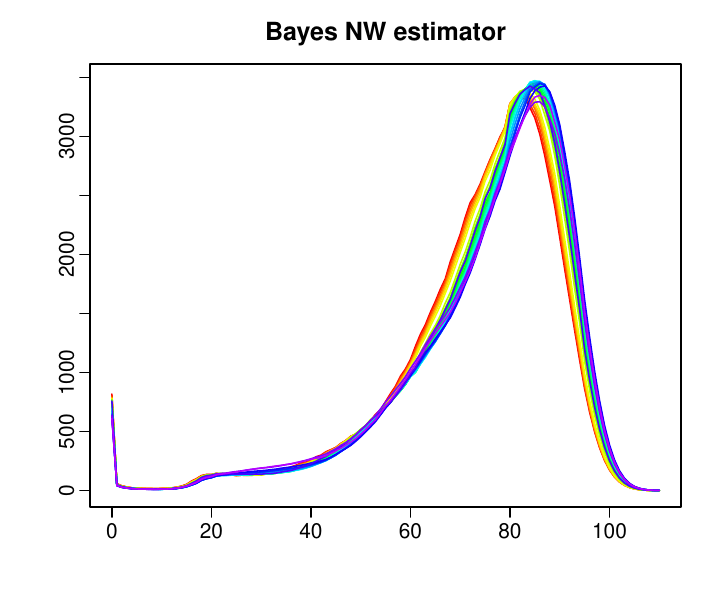}
\quad
\includegraphics[width = 5.6cm]{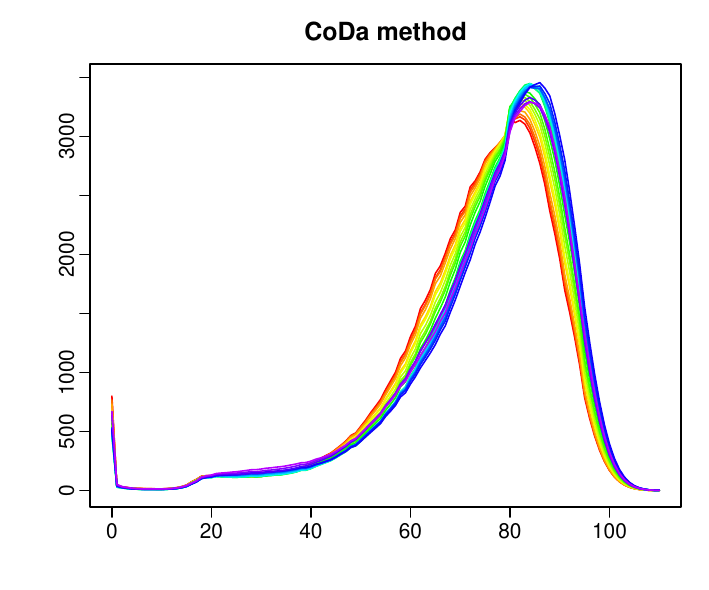}
\\
\includegraphics[width = 5.6cm]{Fig9d}
\quad
\includegraphics[width = 5.6cm]{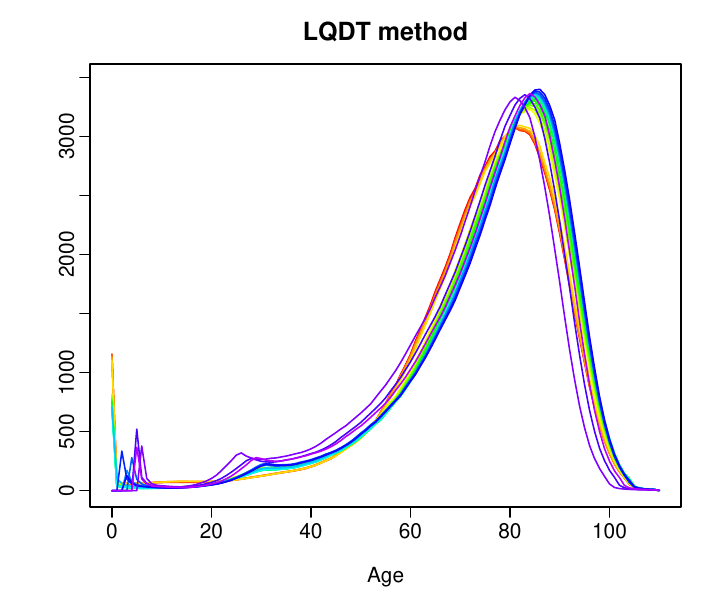}
\quad
\includegraphics[width = 5.6cm]{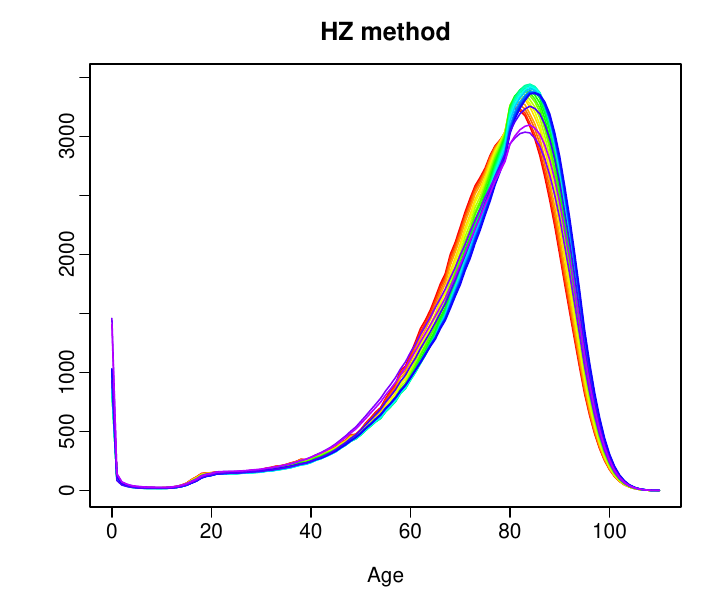}
\caption{\small{Holdout and forecast female and male life-table death counts from 2001 to 2023. The forecast life-table death counts were obtained via the nonparametric density-on-density regression with Bayes NW estimator, CoDa method, LQDT method, and functional principal component regression of \cite{HZ18}, where the optimal bandwidth parameter was selected by the GCV.}}\label{fig:6}
\end{figure}

Regarding the life-table death counts, we split the entire data set into a training sample from 1933 to 2000 and a testing sample from 2001 to 2023. Based on the initial training sample from 1933 to 2000, we compute the one-step-ahead density forecast in 2001 and the forecast error via the KLD. From the data in the training sample from 1933 to 2001, we again compute the one-step-ahead density forecast in 2002 and compute the forecast error. Then, we increase the training sample until the end of the testing samples. In Figure~\ref{fig:6}, we display the density forecasts computed by the nonparametric density-on-density regression with the Bayes NW estimator, CoDa method, LQDT method of \cite{PM16}, and \citeauthor{HZ18}'s \citeyearpar{HZ18} method. Compared with the actual holdout samples, the Bayes NW estimator can adequately capture the local patterns, especially for the male life-table death counts between ages 18 and 40.

\begin{figure}[!htb]
\centering
\includegraphics[width=8.7cm]{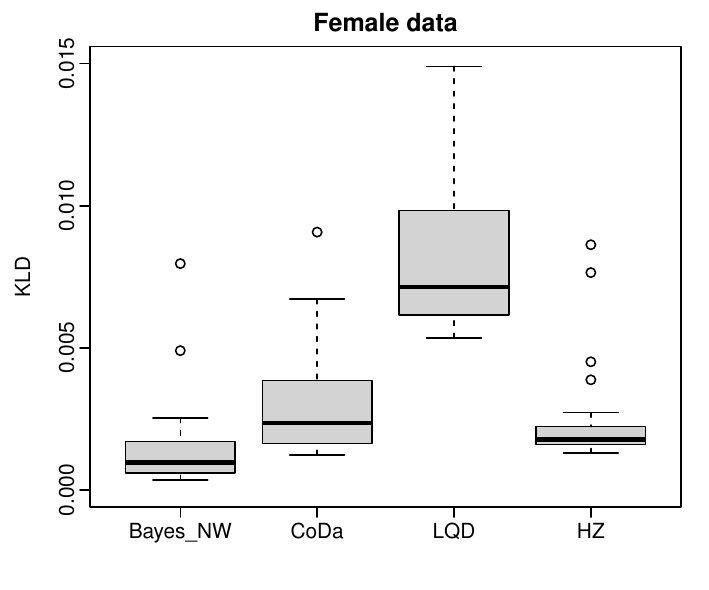}
\quad
\includegraphics[width=8.7cm]{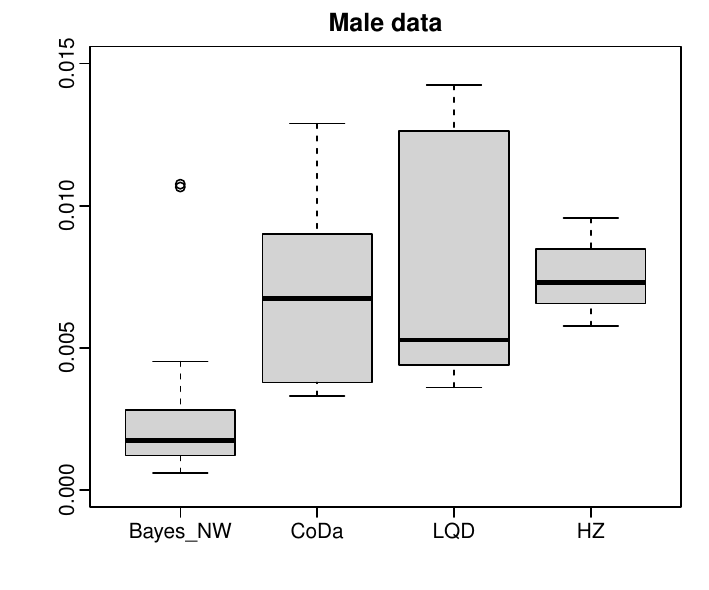}
\caption{\small Forecast accuracy (MSPE ratios) between the nonparametric density-on-density regression using the Bayes NW, CoDa method, LQD method, and functional principal component regression of \cite{HZ18} for forecasting one-step-ahead age-specific life-table death counts from 2001 to 2023 in the USA.}\label{fig:MSPE_LT}
\end{figure}

With the optimal bandwidth parameter selected by the GCV, we iteratively compute the one-step-ahead density forecasts for years from 2001 to 2023, and then compute the overall KLDs in Figure~\ref{fig:MSPE_LT}.

We evaluate and compare the KLD among the Bayes NW, CoDa, LQDT methods, and functional principal component regression of \cite{HZ18}. The KLD of the nonparametric density-on-density regression with the Bayes NW estimator is the smallest among the four methods considered. The inferior performance of the LQDT method is primarily due to the left boundary problem, where there is a comparably higher number of deaths at the infant age. 

\section{Conclusion}\label{sec:5}

We propose a nonparametric density-on-density regression with the Bayes NW estimator to model and forecast density-valued objects. The advantage of the nonparametric density-on-density regression is that the functional time series forecasts within the CoDa do not require a dimension reduction. In turn, there is no loss of information in the proposed method. Simulation and empirical data analyses demonstrate that the nonparametric density-on-density regression achieves good forecast accuracy. 

Through a series of simulation studies, we show that the proposed method is competitive with several existing methods reported in \cite{KMP+19}. Since true densities are often unobserved in empirical applications, we consider a kernel density estimator to estimate densities. The estimation accuracy of such an estimator depends on the type of kernel function and the selection of the bandwidth parameter. For estimating densities, we consider the truncated Gaussian kernel function, and the optimal bandwidth is selected by either Silverman's ROT. For estimating the nonparametric estimator of the density-on-density regression, the optimal bandwidth parameter is selected by the GCV. With the selected bandwidth, we evaluated and compared the one-step-ahead forecast error between the forecast densities constructed based on observed and estimated densities based on a kernel density estimator of the holdout observations using the expanding window approach. For reproducibility, the \Rlogo \ code for forecasting methods and their use in simulation and empirical data analyses are described in \url{https://github.com/hanshang/Bayes_NW}. 



There are at least three ways the current paper can be further extended: First, a future extension is to develop a bandwidth procedure that jointly selects the optimal bandwidth parameters of the kernel density estimator and the Bayes NW estimator. Second, comparing the predictive performance of different methods might not be as fruitful as one might think. It is challenging to determine when a predictive model works because everything is drowned in the incompressible randomness of new observations. The difference between competing methods can often be seen only in the digits of prediction error, even if they encode substantially different pictures of reality. An improved forecast accuracy may be achieved by linearly combining the density forecasts from the nonparametric density-on-density regression and the CoDa method. Finally, we proposed a Bayes NW estimator because of its simplicity. Its extension to the Bayes local linear estimator can produce improved estimation accuracy.

\section*{Acknowledgments}

The authors thank the editors and two reviewers for their detailed comments, which helped us improve our manuscript. We are grateful for the insightful comments received from the participants at the 4\textsuperscript{th} International Conference on Econometrics and Statistics (EcoSta 2021) and the Functional Data Analysis -- A Workshop in Honor of Alois Kneip in 2025. The second author acknowledges financial support from an Australian Research Council Discovery Project (DP230102250) and an Australian Research Council Future Fellowship (FT240100338).

\section*{Conflicts of Interest}

The authors declare no conflicts of interest.

\section*{Data Availability Statement}

The data sets that support the findings are publicly available in the French open data portal \url{https://www.data.gouv.fr/fr/} and the \cite{HMD22}.

\appendixpage
\appendix


\section{Proof of Proposition \ref{prop:dependence}} \label{app:Proposition}

\noindent {\em Proof of (i).} Set $\mathcal{B}_2(I)^k \coloneqq \mathcal{B}_2(I) \times \cdots \times \mathcal{B}_2(I)$ the Cartesian product of $k$ Bayes Hilbert spaces equipped with norm $\|.\|_{\mathcal{B}^k}$ such that $\| (f_1, \ldots, f_k) \|_{\mathcal{B}^k} \coloneqq \sum_{t=1}^k \| f_t \|_\mathcal{B}$ and let $\mathcal{P}(\mathcal{B}_2(I)^k)$ be the space of probability measures on $\mathcal{B}_2(I)^k$ with finite second moment. We define the operator $\mathcal{T}$ such that
\begin{equation*}
\mathcal{T}\, : \, \left\{ \begin{array}{ccl} \mathcal{P}(\mathcal{B}_2(I)^k) & \rightarrow & \mathcal{P}(\mathcal{B}_2(I)^k) \\
\mu & \mapsto & \mathcal{L}\left( (m(f_1)\oplus \epsilon_1, \ldots, m(f_k)\oplus \epsilon_k)  \right),
\end{array}
\right. 
\end{equation*}
where $\mathcal{L}\left( (m(f_1)\oplus \epsilon_1, \ldots, m(f_k)\oplus \epsilon_k)  \right)$ stands for the law of $(m(f_1)\oplus \epsilon_1, \ldots, m(f_k)\oplus \epsilon_k)$ and $f \coloneqq (f_1,\ldots,f_k) \sim \mu$ (that is, $f$ is drawn by $\mu$). Set $g=(g_1, \ldots, g_k) \sim \nu$ with $\nu \in \mathcal{P}(\mathcal{B}_2(I)^k)$; according to the definition of $\|.\|_{\mathcal{B}^k}$ and the operators $\oplus$ and $\ominus$, 
\begin{align*}
\E \| (m(f_1)\oplus \epsilon_1, \ldots, m(f_k)\oplus \epsilon_k) &\ominus_{\mathcal{B}^k} (m(g_1)\oplus \epsilon_1, \ldots, m(g_k)\oplus \epsilon_k) \|_{\mathcal{B}^k}\\
& = \sum_{t=1}^k \E \| (m(f_t)\oplus \epsilon_t) \ominus (m(g_1)\oplus\epsilon_t) \|_\mathcal{B}\\ 
& = \sum_{t=1}^k \E \| m(f_t) \ominus m(g_t)\|_\mathcal{B},
\end{align*}
where $\ominus_{\mathcal{B}^k}$ is the perturbation-subtraction operator in the cartesian product $\mathcal{B}(I)^k$ such that, for any $k$-tuples $(f_1, \ldots, f_k)$ and $(g_1, \ldots, g_k)$, $f \ominus_{\mathcal{B}^k} g \coloneqq (f_1 \ominus g_1, \ldots, f_k \ominus g_k)$. Thanks to the Lipschitz property (H2), there exists $C\in(0,1)$ with, for any $t$,  
$$
\E \| m(f_t) \ominus m(g_t)\|_\mathcal{B} \leq C \, \E \| f_t \ominus g_t\|_\mathcal{B}.
$$ 
As a consequence, there exists $C\in(0,1)$ such that
\begin{align} 
\E \| (m(f_1)\oplus \epsilon_1, \ldots, m(f_k)\oplus \epsilon_k) & \ominus_{\mathcal{B}^k} (m(g_1)\oplus \epsilon_1, \ldots, m(g_k)\oplus \epsilon_k) \|_{\mathcal{B}^k} \nonumber \\ 
& \leq C \, \sum_{t=1}^k \E \| f_t \ominus g_t\|_\mathcal{B} \nonumber \\ 
& \leq C \, \E \| f \ominus_{\mathcal{B}^k} g\|_{\mathcal{B}^k} \label{ineq:lipschitz}
\end{align}
For $(\mu,\nu) \in \mathcal{B}_2(I)^k \times \mathcal{B}_2(I)^k$, let $\Pi(\mu,\nu)$ be the set of probability measures on $\mathcal{B}_2(I)^k \times \mathcal{B}_2(I)^k$ having $\mu$ and $\nu$ as marginals and set
\begin{equation*}
W_2\left( \mu, \nu \right) \coloneqq \inf_{\pi \in \Pi(\mu, \nu)} \int_{\mathcal{B}_2(I)^k\times\mathcal{B}_2(I)^k} \|f \ominus_{\mathcal{B}^k} g \|_{\mathcal{B}^k} \, d \pi(f, g),
\end{equation*}
the Wasserstein 2-distance $W_2\left( \mu, \nu \right)$ between $\mu$ and $\nu$. With the definition of operator $\mathcal{T}$ and~\eqref{ineq:lipschitz},
\begin{equation*}
W_{2}\left( \mathcal{T}(\mu), \mathcal{T}(\nu) \right) \leq C \, W_{2}\left( \mu, \nu \right),
\end{equation*}
The operator $\mathcal{T}$ is a contraction since $C < 1$ (thanks to (H\ref{hypo:regression})) and with Banach's fixed point theorem \citep[see, e.g.,][]{KK11}, there exists a unique probability measure $\mu^*$ such that $\mu^*=\mathcal{T}(\mu^*)$. This means that for any $k$, the law of $(f_1, \ldots, f_k)$ is equal to the law of $(f_2, \ldots, f_{k+1})$. The same developments can be iterated $H$ times so that, for any $k$ and $H$, the law of $(f_1, \ldots, f_k)$ is equal to the law of $(f_{1+H}, \ldots, f_{k+H})$. Conclusion: (H\ref{hypo:regression}) implies the existence of a stationary sequence of random PDFs $\{f_t\}_{t \in \Z}$ that accommodates the nonparametric model $f_{t+1} = m(f_t)\oplus \epsilon_t$.\\ To end the proof of $(i)$, just remark that the error $\epsilon_t$ is a measurable mapping of ($f_t, f_{t+1})$ since $clr(\epsilon_t)= clr(f_{t+1}) - clr\circ m(f_t)$; the $\sigma$-algebra generated by $\epsilon_p, \ldots, \epsilon_q$ (i.e., $\sigma(\epsilon_t: \, p \leq t \leq q)$) is equal to $\sigma(f_t: \, p \leq t \leq q+1)$. Then, for any $H$, the $\rho$-mixing coefficient $\rho_\epsilon(H)$ of $\{\epsilon_t\}_{t\in\Z}$ is equal to $\rho_f(H-1)$, the $\rho$-mixing coefficient of $\{f_t\}_{t\in\Z}$. Consequently, $\{\epsilon_t\}_{t\in\Z}$ inherits the $\rho$-mixing property as soon as $\{f_t\}_{t\in\Z}$ is $\rho$-mixing and satisfies model~\eqref{eq:1}.
\\

\noindent {\em Proof of~(ii).} 

\begin{itemize}
\item {\em About the stationarity of $\{\bX_t\}_{t\in\Z}$}. For measurable sets $B_1, \ldots, B_k \subset \mathcal{I}^n$, 
\begin{equation*}
\text{Prob} \left( \bX_{t_1}\in B_1,\ldots, \bX_{t_k}\in B_k \right) =  \E \left\{ \text{Prob} \left( \bX_{t_1}\in B_1,\ldots, \bX_{t_k}\in B_k | \{f_t\}_{t \in \Z} \right) \right\},
\end{equation*}
and under the conditional probabilistic structure assumed in (H\ref{hypo:dependence}),
\begin{equation*}
\text{Prob} \left( \bX_{t_1}\in B_1,\ldots, \bX_{t_k}\in B_k \right)  =   \int_\Omega  \Pi_{j=1}^k \left\{ \int_{B_j } f_{t_j}(\omega, x)^n dx  \right\} \, d\omega.
\end{equation*}
The right hand side only depends on the law of $f_{t_1}, \ldots, f_{t_k}$ and by stationarity of $\{f_t\}_{t\in\Z}$; it equals the analogous expression with each $t_j$ shifted by any fixed integer $H$: 
\begin{equation*}
\text{Prob} \left( \bX_{t_1}\in B_1,\ldots, \bX_{t_k}\in B_k \right) = \text{Prob} \left( \bX_{t_1+H}\in B_1,\ldots, \bX_{t_k+H}\in B_k \right).
\end{equation*}    
\item {\em About the $\rho$-mixing property of $\{\bX_t\}_{t\in\Z}$}. Set $\mathcal{G}_p^q = \sigma(\bX_t: \, p \leq t \leq q)$ and let $U \in \mathcal{L}^2(\mathcal{G}_{-\infty}^0)$ and $V \in \mathcal{L}^2(\mathcal{G}_{H}^\infty)$ be two rrv with $\var (U)>0$ and $\var (V) > 0$.  Using the same notations introduced to describe the dependence structure (see Section \ref{sec:Prob_and_dep}) and the conditional probabilistic structure of $\{\bX_t\}_{t\in\Z}$ given process $\{f_t\}_{t\in\Z} $ (see (H\ref{hypo:dependence})), 
\begin{eqnarray*}
\E(U V) & = & \E \left\{ \E(U V | \{f_t\}_{t\in\Z}) \right\}\\ 
        & = & \E \left( U_f V_f \right),
\end{eqnarray*}
where $U_f = \E(U  | \{f_t\}_{t\in\Z} )$ and $V_f = \E(V  | \{f_t\}_{t\in\Z} )$ and we have $\cov(U,V) = \cov\left(U_f, V_f\right)$.
By the projection property, $\var(U_f) \leq \var(U)$ and $\var(V_f) \leq \var(V)$ so that 
\begin{eqnarray*}
| \corr(U,V) | & = & \frac{| \cov(U_f, V_f) |}{\sqrt{\var(U) \var(V)}}\\ 
               & \leq & \frac{|\cov(U_f, V_f) |}{\sqrt{\var(U_f) \var(V_f)}}\\
               & \leq & | \corr(U_f,V_f) |.
\end{eqnarray*}
Because $U_f$ (resp. $V_f$) is $\mathcal{F}_{-\infty}^0$-measurable (resp. $\mathcal{F}_{H}^\infty$-measurable), the definition of $\rho_f$ yields $| \corr(U_f,V_f) | \leq \rho_f(H)$, which results in $| \corr(U,V) | \leq \rho_f(H)$. The $\rho$-mixing coefficient $\rho_{\bX}(H)$ of $\{\bX_t\}_{t\in\Z}$ is such that $\rho_{\bX}(H) \leq \rho_f(H)$; $\{\bX_t\}_{t\in\Z}$ is a $\rho$-mixing process as soon as the PDF-valued process $\{f_t\}_{t\in\Z}$ is a $\rho$-mixing process.
\end{itemize}

\section{Proof of main asymptotic results} \label{app:Asym}

\subsection{Preliminaries} \label{app:A}
\noindent\emph{Conditional expectation in $B^2(I)$ and model}. Let $g$ and $h$ two $\mathcal{B}_2(I)$-valued random PDFs; for any $f$ in $\mathcal{B}_2(I)$, the conditional expectation $\E(g | h)$ is such that $
\langle \E( g | h), f \rangle_\mathcal{B} = \E\left(\langle g, f \rangle_\mathcal{B}  | h \right)$ \citep[see, e.g.,][]{Bosq00}. Then, 
$\langle \E(f_{t+1} | f_t), g \rangle_\mathcal{B}  =  \E\left(\langle f_{t+1}, \, g \rangle_\mathcal{B}  | f_t\right) $ and 
 (\ref{eq:1}) entails $$  \langle \E(f_{t+1} | f_t), g \rangle_\mathcal{B}  =  \E\left(\langle m(f_t)\oplus\epsilon , \, g \rangle_\mathcal{B}  | f_t\right).$$ With the property of the inner product, $\langle\cdot, \, \cdot \rangle_\mathcal{B}$, 
\begin{eqnarray*}
\langle \E(f_{t+1} | f_t), g \rangle_\mathcal{B} & = & \E\left(\langle m(f_t), \, g \rangle_\mathcal{B} | f_t \right) + \E\left(\langle\epsilon_t, \, g \rangle_\mathcal{B}  | f_t\right)\\ & = &\langle m(f_t), \, g \rangle_\mathcal{B} + \langle \E(\epsilon_t | f_t), \, g \rangle_\mathcal{B}.
\end{eqnarray*}
According to model~\eqref{eq:1}, $ \E(\epsilon_t | f_t) =  {\bf 0}_\mathcal{B}$ so that, for any $g$ in $\mathcal{B}_2(I)$,  $ \langle \E(f_{t+1} | f_t), g \rangle_\mathcal{B} =  \langle m(f_t), g \rangle_\mathcal{B}$, which is equivalent to $m(f)= \E\left( f_{t+1} | f_t = f \right)$.\\

\subsection{Proof of Theorem~\ref{theorem}} \label{app:B}

Set $\widehat{K}_t \coloneqq K_{\text{reg}} \left(h_{\text{reg}}^{-1} \,  \| \widehat{f}_t   \ominus  f   \|_\mathcal{B} \right)$, $\displaystyle \widehat{A} \coloneqq (  N \, \E  \widehat{K}_1)^{-1}   \odot   \bigoplus_{t=1}^N \left( \widehat{K}_t  \odot \widehat{f}_{t+1}  \right)$, $\displaystyle \widehat{B} \coloneqq (  N \, \E  \widehat{K}_1)^{-1} \sum^N_{t=1} \, \widehat{K}_t$; the proof is based on the decomposition 
\begin{equation*} 
\widehat{m}_N(f)  \ominus m(f) = \widehat{B}^{-1} \odot \left\{ \widehat{A} - \E \widehat{A} \right\} \oplus   \widehat{B}^{-1} \odot \left\{ \E \widehat{A} \ominus m(f) \right\} \oplus \left( \widehat{B}^{-1} - 1  \right)  \odot m(f),
\end{equation*}
which results in 
\begin{equation} \label{eq:theo_decomposition}
\displaystyle \| \widehat{m}_N(f)  \ominus m(f) \|_\mathcal{B}  \leq  
\widehat{B}^{-1}  \left\{ \underbrace{ \|   \widehat{A}   \ominus    \E \widehat{A}  \|_\mathcal{B} }_{Q_1} +  \underbrace{\|   \E \widehat{A}   \ominus   m(f) \|_\mathcal{B} }_{Q_2} \right\} +  \widehat{B}^{-1} \underbrace{ | 1  -    \widehat{B} | }_{Q_3}. 
\end{equation}
Before going on, let us point out useful features of the Bayes space $\mathcal{B}_2(I)$ that are systematically used in the proofs. First,  the Bayes space of bounded PDFs is a separable Hilbert space; there exists an orthonormal basis $\{e_j\}_{j \geq 1}$ such that, for any $g$ in $\mathcal{B}_2(I)$,   $g = \sum_{j\geq 1}     \langle g,  e_j \rangle_\mathcal{B}^2   $.  Second, the standard properties of the norms naturally apply in $\mathcal{B}_2(I)$; for instance, for any $u_1$, $u_2$ in $\R$ and any $g_1$, $g_2$ in $\mathcal{B}_2(I)$, $\|  u_1  \odot   g_1  \oplus  u_2  \odot  g_2  \|_\mathcal{B}  \leq     | u_1 |  \|  g_1   \|_\mathcal{B}    +   |  u_2  |  \|   g_2  \|_\mathcal{B}$. Third, with the assumptions on the kernel function  (\prettyref{hypo:kernel}), it exists $C>0$ and $C'>0$ such that 
$$
C\, 1_{[0,1]}\left(h_{\text{reg}}^{-1} \,  \|f_{1} \ominus  f\|_\mathcal{B} \right) \leq K_1 \leq C'\,  1_{[0,1]}\left(h_{\text{reg}}^{-1} \,  \|f_{1} \ominus  f \|_\mathcal{B} \right),
$$
where $K_1 \coloneqq K_{\text{reg}} \left(h_{\text{reg}}^{-1} \,  \| f_t   \ominus  f   \|_\mathcal{B} \right)$. Then,
\begin{equation} \label{eq:theo_sbp1}
C \, \pi_f(h_{\text{reg}})  \leq   \E \, K_1   \leq   C'  \pi_f(h_{\text{reg}}).
\end{equation}
The same result holds when $f_1$ is replaced with its estimator $\widehat{f}_1$:
\begin{equation} \label{eq:theo_sbp2}
C \, \widehat{\pi}_f(h_{\text{reg}})  \leq   \E \, \widehat{K}_1   \leq   C'  \widehat{\pi}_f(h_{\text{reg}}),
\end{equation}
with $  \widehat{\pi}_f(h_{\text{reg}})  \coloneqq   P(\|\widehat{f}_1  \ominus  f   \|_\mathcal{B}  \leq  h_{\text{reg}})$. 

\begin{rem}
It is easy to see that similar arguments provide
\begin{equation} \label{eq:theo_sbp3}
C^2 \, \pi_f(h_{\text{reg}})  \leq   \E \,  K_1^2   \leq   C'^2  \pi_f(h_{\text{reg}}).
\end{equation}
The same result holds when $f_1$ is replaced with its estimator $\widehat{f}_1$:
\begin{equation} \label{eq:theo_sbp4}
C^2 \, \widehat{\pi}_f(h_{\text{reg}})  \leq   \E \, \widehat{K}_1^2   \leq   C'^2  \widehat{\pi}_f(h_{\text{reg}}).
\end{equation}
\end{rem}
From now on, $C$ and $C'$ denote two generic strictly positive constants. \\ 
~\\ 
\emph{Focus on $Q_1$}. According to the definition of $Q_1$ in (\ref{eq:theo_decomposition}),  $\mathcal{B}_2(I)$, $Q_1^2 = \sum_{j\geq 1}  \langle   \widehat{A}   \ominus    \E \widehat{A} ,  e_j   \rangle_\mathcal{B}^2$:
\begin{eqnarray*}
Q_{1}^{2} & = & (N \, \E  \widehat{K}_1)^{-2} \sum_{j\geq 1} \langle \bigoplus_{t=1}^N \left\{ \left(\widehat{K}_t  \odot \widehat{f}_{t+1} \right) \ominus \E \left( \widehat{K}_t  \odot \widehat{f}_{t+1}  \right) \right\} , e_j \rangle_\mathcal{B}^2\\
& = & (  N \, \E  \widehat{K}_1)^{-2} \sum_{t = 1}^N \sum_{j\geq 1} \left\{ \widehat{K}_t \langle \widehat{f}_{t+1}, e_j \rangle_\mathcal{B} - \E \left( \widehat{K}_t \langle \widehat{f}_{t+1}, e_j \rangle_\mathcal{B} \right) \right\}^2 \\ 
& & + (  N \, \E  \widehat{K}_1)^{-2} \sum_{s = 1}^N  \sum_{t = 1}^N \sum_{j\geq 1} \left( \widehat{K}_s \langle \widehat{f}_{s+1}, e_j \rangle_\mathcal{B} - \E \widehat{K}_s\langle \widehat{f}_{s+1}, e_j \rangle_\mathcal{B} \right) \left( \widehat{K}_t \langle \widehat{f}_{t+1}, e_j \rangle_\mathcal{B} - \E \widehat{K}_t \langle \widehat{f}_{t+1}, e_j \rangle_\mathcal{B} \right).
\end{eqnarray*}
Let $u_N$ be a sequence tending to infinity with $N$ and set $\text{cov}_{s,t,j} \coloneqq \text{Cov}  \left(  \widehat{K}_s  \langle   \widehat{f}_{s+1}  ,  \, e_j   \rangle_\mathcal{B}   , \,   \widehat{K}_t  \langle   \widehat{f}_{t+1}  ,  e_j   \rangle_\mathcal{B}     \right)$: 
\begin{equation}\label{eq:asymp_theo1}
\E Q_1^2 = Q_{1,1} + Q_{1,2} + Q_{1,3}
\end{equation}
with 
\begin{align*}
Q_{1,1}     &\coloneqq   (  N \, \E  \widehat{K}_1)^{-2}  \sum_{t=1}^N  \sum_{j\geq 1}  \text{Var}  \left(  \widehat{K}_t  \langle   \widehat{f}_{t+1}  ,  e_j   \rangle_\mathcal{B}     \right), \\
 Q_{1,2}     &\coloneqq  (  N \, \E  \widehat{K}_1)^{-2}   \underset{|s - t| < u_N}{\sum_{s=1}^N  \sum_{t=1}^N}  \sum_{j\geq 1}  \text{cov}_{s,t,j}, \\
Q_{1,3}     &\coloneqq  (  N \, \E  \widehat{K}_1)^{-2}   \underset{|s - t| > u_N}{\sum_{s=1}^N  \sum_{t=1}^N  }  \sum_{j\geq 1}  \text{cov}_{s,t,j}.
\end{align*}
\begin{itemize}
\item About $Q_{1,1}$. 
\begin{eqnarray*}
\sum_{j\geq 1} \text{Var}  \left(  \widehat{K}_t  \langle   \widehat{f}_{t+1}  ,  e_j   \rangle_\mathcal{B}     \right) &\leq & \sum_{j\geq 1}  \E \left(  \widehat{K}_t^2  \langle   \widehat{f}_{t+1}  ,  \, e_j   \rangle_\mathcal{B}^2  \right) \\ & \leq & \E \left(  \widehat{K}_1^2  \|   \widehat{f}_{2} \|_\mathcal{B}^2  \right)\\ 
& \leq &  \E \left(  \widehat{K}_1^2  \|   \widehat{f}_{2} - f\|_\mathcal{B}^2  \right) + \E \left(  \widehat{K}_1^2  \| f \|_\mathcal{B}^2  \right),
\end{eqnarray*}
where the second inequality comes from the stationarity property. As the estimator (\ref{def:regression_estimator_practice}) only involves $\widehat{f}_t$'s such that $ \|   \widehat{f}_{t} - f\|_\mathcal{B}^2 \leq h_{\text{reg}}$ and thanks to~\eqref{eq:theo_sbp4}, 
        \begin{eqnarray*}
            \sum_{j\geq 1} \text{Var}  \left(  \widehat{K}_t  \langle   \widehat{f}_{t+1}  ,  e_j   \rangle_\mathcal{B}     \right) & \leq & (h_{\text{reg}} + C) \, \E \widehat{K}_1^2 \\
            & = & O_P\left\{\widehat{\pi}_f(h_{\text{reg}})\right\}.
        \end{eqnarray*}
        Because $Q_{1,1}     =  N^{-1} \, ( \E  \widehat{K}_1)^{-2} \sum_{j\geq 1} \text{Var}  \left(  \widehat{K}_t  \langle   \widehat{f}_{t+1}  ,  e_j   \rangle_\mathcal{B}     \right) $ and $( \E  \widehat{K}_1)^{-2} = O \left(  \left\{  \widehat{\pi}_f(h_{\text{reg}})\right\}^{-2}   \right)$,
        \begin{equation} \label{eq:theo_Q11}
            Q_{1,1}     =   O \left(  \left\{  N  \, \widehat{\pi}_f(h_{\text{reg}})\right\}^{-1}   \right).
        \end{equation}
    \item 
        About $Q_{1,2}$. With the Cauchy-Schwartz inequality and the dependence structure, 
        \begin{equation*}
            \E \left(  \widehat{K}_s  \langle   \widehat{f}_{s+1}  ,  \, e_j   \rangle_\mathcal{B}  \,   \widehat{K}_t  \langle   \widehat{f}_{t+1}  ,  e_j   \rangle_\mathcal{B}     \right) \leq  \E  \left( \widehat{K}_1^2  \langle   \widehat{f}_2  ,  e_j   \rangle_\mathcal{B}^2  \right),
        \end{equation*}
        and one has
        \begin{equation*}
            \underset{|s - t| < u_N}{\sum_{s=1}^N  \sum_{t=1}^N} \sum_{j\geq 1} |  \text{cov}_{s,t,j}  |   \leq 2 N \, u_N \,  \E  \left( \widehat{K}_1^2  \|  \widehat{f}_2  \|_\mathcal{B}^2  \right).
        \end{equation*}
        Similarly to the study of $Q_{1,1}$, $\E  \left( \widehat{K}_1^2  \|  \widehat{f}_2  \|_\mathcal{B}^2  \right) = O_P\left\{\widehat{\pi}_f(h_{\text{reg}})\right\}$, which results in 
        \begin{equation}\label{eq:theo_Q12}
            Q_{1,2} = O \left( u_N \left\{  N  \, \widehat{\pi}_f(h_{\text{reg}})   \right\}^{-1}  \right).
        \end{equation}
        
    \item About $Q_{1,3}$. 
        \begin{eqnarray*}
            |\text{cov}_{s,t,j}| & = &  \left|\text{Corr}  \left(  \widehat{K}_s  \langle   \widehat{f}_{s+1}  ,  \, e_j   \rangle_\mathcal{B}   , \,   \widehat{K}_t  \langle   \widehat{f}_{t+1}  ,  e_j   \rangle_\mathcal{B}     \right) \right |\, \text{Var}\left( \widehat{K}_s  \langle   \widehat{f}_{s+1}  ,  \, e_j   \rangle_\mathcal{B} \right)\\
            & \leq & \rho_f(|s-t|) \, \E \left( \widehat{K}_s^2  \| \widehat{f}_{s+1} \|_\mathcal{B}^2 \right),
        \end{eqnarray*}
            where the last inequality is a consequence of the definition of the $\rho$-mixing coefficient (\ref{def:mixing_coefficient}). Again, $\E  \left( \widehat{K}_1^2  \|  \widehat{f}_2  \|_\mathcal{B}^2  \right) = O_P\left\{\widehat{\pi}_f(h_{\text{reg}})\right\}$; because $|s-t|> u_N$ and thanks to (\prettyref{hypo:mixing}), 
        \begin{equation}\label{eq:theo_Q13}
                Q_{1,3}   =   O_P  \left( \left\{  u_N^a  \, \widehat{\pi}_f(h_{\text{reg}})   \right\}^{-1}  \right).
        \end{equation}
\end{itemize}
Set $u_N  \coloneqq  N^{1/(a+1)}$, the quantity balancing  $Q_{1,2}$ and $Q_{1,3}$;  from (\ref{eq:asymp_theo1})-(\ref{eq:theo_Q13}), one has
\begin{equation} \label{eq:theo_Q1}
    Q_1   =   O_P  \left( \left\{  N^{a/(a+1)}  \, \widehat{\pi}_f(h_{\text{reg}})   \right\}^{-1/2}  \right).
\end{equation}
 ~\\ 
\emph{Focus on $Q_2$}. Taking into account the stationarity property, 
\begin{eqnarray} \nonumber
    Q_2 & = & \underbrace{( \E  \widehat{K}_1)^{-1}    \E  \left(  \widehat{K}_1  \|  \widehat{f}_2  \ominus   f_2  \|_\mathcal{B}  \right)}_{Q_{2.1}} + \underbrace{(   \E   \widehat{K}_1)^{-1}     \E  \left(    \widehat{K}_1    \| m(f_1)  \ominus  m(f)  \|_\mathcal{B}  \right)}_{Q_{2.2}}\\ & &  +  \underbrace{(   \E   \widehat{K}_1)^{-1}    \left\| \E  \left(    \widehat{K}_1  \odot  \epsilon_1   \right) \right\|_\mathcal{B} }_{Q_{2.3}}. \label{eq:asymp_theo2}
\end{eqnarray}
Cauchy-Schwartz inequality with  (\prettyref{hypo:estdensity}) and (\ref{eq:theo_sbp4}) entails  $  \E  \left(  \widehat{K}_1  \|  \widehat{f}_2 \ominus   f_2  \|_\mathcal{B}  \right)    =     O   \left(   \widehat{\pi}_f(h_{\text{reg}})^{1/2}    \delta_N      \right)   $; use again (\prettyref{hypo:kernel}) to get    
\begin{equation}\label{eq:theo_Q21}
    Q_{2,1}  =   O  \left(   \widehat{\pi}_f(h_{\text{reg}})^{-1/2}    \delta_{N} \right).
\end{equation}
With the Lipschitz property of the regression operator $m$ (see (\prettyref{hypo:regression})),   $$   \| m(f_1)  \ominus  m(f)  \|_\mathcal{B}   \leq  C    \|  f_1  \ominus  f  \|_\mathcal{B}  .$$ Because we only retain $f_t$'s such that $K_t>0$, $   \| m(f_1)  \ominus  m(f)  \|_\mathcal{B}   \leq  C  \, h_{\text{reg}}$  and   
\begin{equation}\label{eq:theo_Q22}
    Q_{2,2}  =   O  \left(  h_{\text{reg}}  \right).
\end{equation}
\noindent $ \E  \left(   \widehat{K}_1  \odot  \epsilon_1   \right)   =   \E \left\{  \E  \left(   \widehat{K}_1  \odot  \epsilon_1  |  f_1  \right)    \right\}  $; $\forall g \in \mathcal{B}_2(I)$, 
\begin{eqnarray*}
    \langle  \E  \left(   \widehat{K}_1  \odot  \epsilon_1  |  f_1  \right)   ,  \, g  \rangle_\mathcal{B} & = &  \widehat{K}_1 \,   \langle  \E  \left(    \epsilon_1  |  f_1  \right)   ,  \, g  \rangle_\mathcal{B} \\ & =  & \widehat{K}_1 \,   \langle {\bf 0}_\mathcal{B}   ,  \, g  \rangle_\mathcal{B}\\ &  = & 0.
\end{eqnarray*}
Conclusion:  for any $g$ in $\mathcal{B}_2(I)$,   $  \langle  \E  \left(   \widehat{K}_1  \odot  \epsilon_1  |  f_1  \right)   ,  \, g  \rangle_\mathcal{B}  =   0$, which results in   $   Q_{2,3}   =  0$.

With this last result, (\ref{eq:theo_sbp4}), and (\ref{eq:theo_Q21})-(\ref{eq:theo_Q22}), 
\begin{equation} \label{eq:theo_Q2}
    Q_2  =    O(h_{\text{reg}})    +   O  \left(   \widehat{\pi}_f(h_{\text{reg}})^{-1/2}    \delta_N      \right).
\end{equation}
 ~\\ 
\emph{Remaining terms and summary}. $Q_3$ is a particular case of $Q_1$ when $\widehat{f}_{t+1}$ is replaced with $1$;  $  \widehat{B}  =   1  +  (   \widehat{B}  -  1  )  $ entails  $  \widehat{B}   \leq   1  +  Q_3$ and with (\ref{eq:theo_Q1}),
\begin{equation} \label{eq:theo_B}
    \widehat{B}     =   1   +   O_P  \left( \left\{  N^{a/(a+1)}  \, \widehat{\pi}_f(h_{\text{reg}})   \right\}^{-1/2}  \right).
\end{equation}
Then,  the decomposition (\ref{eq:theo_decomposition}) with (\ref{eq:theo_Q1}), (\ref{eq:theo_Q2}) and (\ref{eq:theo_B}) results in
\begin{equation}  \label{eq:th1.1}
 \displaystyle \| \widehat{m}_N(f)  \ominus m(f) \|_\mathcal{B}  =    O(h_{\text{reg}})    +   O  \left(   \widehat{\pi}_f(h_{\text{reg}})^{-1/2}    \delta_N      \right)   +    O_P  \left( \left\{  N^{a/(a+1)}  \, \widehat{\pi}_f(h_{\text{reg}})   \right\}^{-1/2}  \right). 
\end{equation}

\noindent \emph{Comparison between $\widehat{\pi}_f(h_{\text{reg}})$  and $\pi_f(h_{\text{reg}})$}. We now compare the asymptotic behaviour of $\pi_f(h)$ with that of $\widehat{\pi}_f(h)$. 
\begin{itemize}
\item Thanks to Markov's inequality, for any $C>0$, there exists $M>0$,
    \begin{equation} \label{eq:asymp_theo3}
    P\left( \|  \widehat{f}_1  \ominus  f_1 \|_\mathcal{B} > C \, \delta_N^{1-b} \right) \leq M \, \delta_N^{-2(1-b)} \, \E \left( \|  \widehat{f}_1  \ominus  f_1 \|_\mathcal{B}^2 \right).
    \end{equation}
    According to (\prettyref{hypo:estdensity}), it exists a sequence $\delta_{N}$ tending to 0 as $N$ goes to infinity such that $\E \left( \|  \widehat{f}_1  \ominus  f_1 \|_\mathcal{B}^2 \right) = O(\delta_N^2)$. Then, for any $C>0$, there exists $M>0$,
    \begin{equation*}
    P\left( \|  \widehat{f}_1  \ominus  f_1 \|_\mathcal{B} > C \, \delta_N^{1-b} \right) \leq M \, \delta_N^{2b}.
    \end{equation*}
    Let 
    $$
    \widehat{\mathcal{A}} \coloneqq \left\{  \|   \widehat{f}_1  \ominus    f  \|_\mathcal{B} < h_{\text{reg}} \right\} \text{ and } \widehat{\mathcal{A}}_0 \coloneqq \left\{  \|   \widehat{f}_1  \ominus    f_1  \|_\mathcal{B} > C \, \delta_N^{1-b} \right\}
    $$ 
    be two events; $\widehat{\mathcal{A}} = \widehat{\mathcal{A}}_1 \cup \widehat{\mathcal{A}}_2$ with $\widehat{\mathcal{A}}_1 \coloneqq \widehat{\mathcal{A}} \cap \widehat{\mathcal{A}}_0$ and $\widehat{\mathcal{A}}_2 \coloneqq \widehat{\mathcal{A}} \cap \overline{\widehat{\mathcal{A}}_0}$. (\ref{eq:asymp_theo3}) entails 
    \begin{equation} \label{eq:asymp_theo4}
        P(\widehat{\mathcal{A}}_1) \leq P(\widehat{\mathcal{A}}_0) \leq M \, \delta_N^{2b}.
    \end{equation}
    By definition of $\widehat{\mathcal{A}}_2$, $$\widehat{\mathcal{A}}_2 = \left\{  \|   \widehat{f}_1  \ominus    f  \|_\mathcal{B} < h_{\text{reg}} \right\} \cap \left\{  \|   \widehat{f}_1  \ominus    f_1  \|_\mathcal{B} \leq C \, \delta_N^{1-b} \right\} $$  and because $\|   f    \ominus    f_1   \|_\mathcal{B} <   \|    f    \ominus    \widehat{f}_1  \|_\mathcal{B} +  \|  \widehat{f}_1   \ominus   f_1  \|_\mathcal{B}$, $ \widehat{\mathcal{A}}_2 \subset \left\{  \|   f    \ominus    f_1   \|_\mathcal{B} < h_{\text{reg}} + C  \, \delta_N^{1-b} \right\} $, which results in
    \begin{equation} \label{eq:asymp_theo5}
        P(\widehat{\mathcal{A}}_2) \leq \pi_f(h + C \, \delta_N^{1-b}).
    \end{equation}
    With (\ref{eq:asymp_theo4})-(\ref{eq:asymp_theo5}) and $P(\widehat{\mathcal{A}}) \leq P(\widehat{\mathcal{A}}_1) + P(\widehat{\mathcal{A}}_2)$,
    \begin{equation} \label{eq:asymp_theo5bis}
    P(\widehat{\mathcal{A}}) \leq \pi_f(h_{\text{reg}} + C \, \delta_N^{1-b}) + M \, \delta_N^{2b}.
    \end{equation}
    Thanks to (\prettyref{hypo:sbp}) and (\prettyref{hypo:asymptotics}), $\delta_N^{2b} = o\left\{\pi_f(h_{\text{reg}})\right\}$, $\delta_N^{1-b} = o(h_{\text{reg}})$ and $\pi_f\left\{ h_{\text{reg}} + o(h_{\text{reg}})\right\}$; there exists $C>0$ such that 
    \begin{equation} \label{eq:asymp_theo6}
        \widehat{\pi}_f(h_{\text{reg}}) \leq C \, \pi_f(h_{\text{reg}}).
    \end{equation}
\item Set $\widehat{\mathcal{A}}_{\delta_N} \coloneqq \left\{  \|  \widehat{f}_1  \ominus  f  \|_\mathcal{B} < h_{\text{reg}} - C \, \delta_N^{1-b} \right\}$,  $\widehat{\mathcal{A}}_{\delta_N, 1} \coloneqq \widehat{\mathcal{A}}_{\delta_N} \cap \widehat{\mathcal{A}}_0$, and $\widehat{\mathcal{A}}_{\delta_N, 2} \coloneqq \widehat{\mathcal{A}}_{\delta_N} \cap \overline{\widehat{\mathcal{A}}_0}$. Because $ \widehat{\mathcal{A}}_{\delta_N} = \widehat{\mathcal{A}}_{\delta_N, 1} \cup \widehat{\mathcal{A}}_{\delta_N, 2} $,    
$$
P(\widehat{\mathcal{A}}_{\delta_N}) \leq P(\widehat{\mathcal{A}}_{\delta_N, 1}) + P(\widehat{\mathcal{A}}_{\delta_N, 2}). 
$$
With the definitions of $\widehat{\mathcal{A}}_{\delta_N, 1}$ and $\widehat{\mathcal{A}}_{\delta_N, 2}$, $P(\widehat{\mathcal{A}}_{\delta_N, 1}) \leq P(\widehat{\mathcal{A}}_0) $ and $P(\widehat{\mathcal{A}}_{\delta_N, 2}) \leq P(\widehat{\mathcal{A}}_{\delta_N}) \leq P(\widehat{\mathcal{A}})$; this entails
$$
P(\widehat{\mathcal{A}}_{\delta_N}) \leq P(\widehat{\mathcal{A}}_0) + P(\widehat{\mathcal{A}}).
$$
According to (\ref{eq:asymp_theo4}), $\pi_f(h_{\text{reg}} - C \, \delta_N^{1-b}) - M \, \delta_N^{2b} \leq P(\widehat{\mathcal{A}})$ and thanks to (\prettyref{hypo:sbp})-(\prettyref{hypo:asymptotics}), there exists $C>0$ such that 
\begin{equation*}
    C\, \pi_f(h_{\text{reg}}) \leq \widehat{\pi}_f(h_{\text{reg}}),
\end{equation*}
so that it exist $C>0$ and $C'>0$,
\begin{equation*}
    C\, \pi_f(h_{\text{reg}}) \leq \widehat{\pi}_f(h_{\text{reg}})  \leq C' \, \pi_f(h_{\text{reg}}).
\end{equation*}

\end{itemize}
To end the proof of {\sc Theorem} \ref{theorem}, just replace $\widehat{\pi}_f(h_{\text{reg}})$ with $\pi_f(h_{\text{reg}})$ into (\ref{eq:th1.1}).

\subsection{Proof of Theorem~\ref{theo2:rate}} \label{app:proof_theo2}
{\em Proof of $(i)$}.
\begin{itemize}
\item We first proof that $\E \| \widehat{f}_t \ominus f_t \|_\mathcal{B}^2 \leq 4 \, c^{-2} \E\| \widehat{f}_t - f_t \|^2 $. With the isometric feature of the clr transformation, $\| \widehat{f}_t \ominus f_t \|_\mathcal{B}  =  \| \text{clr}(\widehat{f}_t) - \text{clr}(f_t)  \|$. Thanks to the definition of $\text{clr}$, $\text{clr}(\widehat{f}_t) - \text{clr}(f_t) = \ln \widehat{f}_t - \ln f_t - (b-a)^{-1} \int_I ( \ln \widehat{f}_t - \ln f_t)$:
\begin{eqnarray*}
\| \text{clr}(\widehat{f}_t) - \text{clr}(f_t) \|  & \leq &   \| \ln \widehat{f}_t - \ln f_t \| + \int_I | \log \widehat{f}_t - \log f_t |\\
& \leq & c^{-1} (\| \widehat{f}_t - f_t \| + \int_I | \widehat{f}_t - f_t |),
\end{eqnarray*} 
where the last inequality uses the Lipschitz property of the logarithm on the interval $[c, +\infty)$. With the Jensen inequality, $\int_I | \widehat{f}_t - f_t | \leq \| \widehat{f}_t - f_t \|$ so that $\| \widehat{f}_t \ominus f_t \|_\mathcal{B} \leq 4 \, c^{-2} \| \widehat{f}_t - f_t \|^2$, which results in 
\begin{equation} \label{eq:asymp_kde1}
E \| \widehat{f}_t \ominus f_t \|_\mathcal{B}^2 \leq 4 \, c^{-2}\, \E\| \widehat{f}_t - f_t \|^2.
\end{equation} 
\item Focus on $\E\| \widehat{f}_t - f_t \|^2$. 
\begin{eqnarray*}
\E\| \widehat{f}_t - f_t \|^2 & = & \E \left\{ \E \, \left(\| \widehat{f}_t - f_t \|^2 | f_t \right) \right\} \\ \nonumber
& = & \E \int_I \E\left\{ \left( \widehat{f}_t(x) - f_t(x) \right)^2 | f_t \right\} dx.
\end{eqnarray*}
Conditioning on $f_t$ remains to consider a deterministic $f_t(\omega,x)$ for a given $\omega\in\Omega$; let $\text{Bias}(f_t, \omega, x) = \E \left( \widehat{f}_t(x) | f_t \right) - f_t(\omega, x)$ be the conditional bias:
\begin{equation} \label{eq:asymp_kde1bis}
\E \, \| \widehat{f}_t - f_t \|^2  
= \E \, \int_I \text{Bias}(f_t,\omega,x)^2 dx + \E \, \int_I \var\left(  \widehat{f}_t(x)| f_t \right) \, dx. 
\end{equation}
According to the probabilistic structure assumed with~\eqref{hypo:dependence}, given $\omega \in \Omega$, $f_t(\omega,)$ is the PDF of $X_{t,1}$ and $X_{t,1},\ldots, X_{t,n}$ are iid rrv. Then, for each $t$, one can use standard developments \citep[see, e.g.,][]{Rosenblatt71} to get the asymptotic behaviour of the kernel density estimator~\eqref{eq:density_est} conditionally to $f_t$. According to the definition~\eqref{eq:density_est} of our kernel density estimator, 
\begin{eqnarray} \nonumber
\E\left( \widehat{f}_t(x) | f_t \right) & = & \frac{1}{h_t} \int_I K_{\text{kde}}\left( \frac{x - u}{h_t} \right) \, f_t(\omega, u) \, du \\ \nonumber
& = &  \int_{-1}^1 K_{\text{kde}}\left( z \right) \, f_t(\omega, x - h_t z) \, dz \\
& = & f_t(\omega,x) + 0.5 \, h_t^2  \left\{ \mu_2 \, f_t''(\omega,x) + T(\omega, x,\theta,h_t) \right\}. \label{eq:asymp_kde2}
\end{eqnarray}
where $\mu_2 = \int_{-1}^1z^2\, K_{\text{kde}}(z)\, dz$ and  $T(\omega, x,\theta,h_t) = \int_{-1}^1 z^2 K_{\text{kde}}(z)\{f_t''(\omega,x-\theta z h_t) - f_t''(\omega,x)\}dz$ with $\theta\in(0,1)$. The features of the kernel $K_{\text{kde}}$ (i.e. $\int_{-1}^1 K_{\text{kde}}(z)\,dz =1$ and $\int_{-1}^1 z \, K_{\text{kde}}(z)\, dz =0$) and the Taylor expansion of $f_t(\omega, .)$ with (H\ref{hypo:densityderiv}) entails the last equality (\ref{eq:asymp_kde2}). Set $y=x-\theta z h_t$; $\theta\in(0,1)$ and $|z|<1$ imply that $|y-x|<h_t$. Then, 
$$
T(\omega, x,\theta,h_t) = O \left( \sup_{\{|r-s|< h_t\}} |f_t''(\omega,r) - f_t''(\omega,s)|\right). 
$$
Thanks to (H\ref{hypo:densityasym}), it exists $C>0$ such that 
\begin{equation}\label{eq:asymp_kde3} 
T(\omega, x,\theta,h_t) = O \left\{U_N(\omega)\right\},
\end{equation} 
where $U_N(\omega) =  \sup_{\{|r-s|< C N^{-q/5}\}} |f_t''(\omega,r) - f_t''(\omega,s)|$ and with (\ref{eq:asymp_kde2})-(\ref{eq:asymp_kde3}),
\begin{equation}\label{eq:asymp_kde3bis}
\text{Bias}(f_t,\omega,x) = N^{-2q/5} \left\{ O( f_t''(\omega,x) ) + \,O\left(U_N(\omega)\right) \right\}.
\end{equation}
According to (H\ref{hypo:densityderiv}), $\E \| f_t'' \|^2 < \infty$, which entails that 
$$
\E\left( \int_I \text{Bias}(f_t,.,x)^2 dx \right) = N^{-4q/5} \,  \left\{ O \left( 1  \right) + O\left( \sqrt{\E\,U_N^2} \right)\right\}.
$$
The continuity of $f_t''(\omega,.)$ over the compact set $I$ implies the uniform continuity of $f_t''(\omega,.)$ and  $U_N(\omega)$ tends to 0 when $N$ tends to $\infty$. By dominated convergence theorem, $\E \, (U_N^2) = \int_\Omega U_N(\omega)^2\, dP(\omega) \rightarrow 0 \text{ when }N\rightarrow \infty$. Then,
\begin{equation} \label{eq:asymp_kde4}
\E\left( \int_I \text{Bias}(f_t,.,x)^2 dx \right) = O\left(N^{-4q/5}\right),
\end{equation} 
which is the asymptotic behaviour of the unconditional bias term.
    
We now focus on the conditional variance term: 
\begin{eqnarray} \nonumber
\var\left( \widehat{f}_t(x) | f_t  \right) & = &  \frac{1}{n\, h_t^2}\var\left\{ K_{\text{kde}}\left( \frac{x - X_{t,1}}{h_t} \right) \, | \, f_t \right\}\\ \nonumber
& = & \frac{1}{n\, h_t^2} \, \E \left\{K_{\text{kde}}\left( \frac{x - X_{t,1}}{h_t} \right)^2 \, | \, f_t \right\}   - \frac{1}{n} \left\{ \E\left( \widehat{f}_t(x) | f_t \right)\right\}^2 \\ \nonumber
& = & \frac{1}{n h_t} \int_{-1}^1 K_{\text{kde}}(z)^2 f_t(\omega, x-h_t z) \, dz - \frac{1}{n} \left\{ \E\left( \widehat{f}_t(x) | f_t \right)\right\}^2\\ 
& = & T_1(f_t, \omega,x) - T_2(f_t, \omega,x), \label{eq:asymp_kde3a}
\end{eqnarray}
where 
$$
T_1(\omega,x) =(n h_t)^{-1}\int_{-1}^1 K_{\text{kde}}(z)^2 f_t(\omega, x-h_t z) \, dz
$$
and 
$$
T_2(\omega,x) = n^{-1} \left\{ \text{Bias}(f_t,\omega,x)^2 - 2 f_t(\omega,x) \text{Bias}(f_t,\omega,x) + f_t(\omega,x)^2\right\}.
$$
With similar arguments as those used to get the asymptotic behaviour of the conditional pointwise bias term (\ref{eq:asymp_kde3bis}), it exists $C>0$, 
\begin{equation*}
T_1(\omega,x)  =  N^{-4q/5} \left\{  O( f_t(\omega,x) )\, + \, N^{-q/5}\,  O( f_t''(\omega,x) ) \,+ \, N^{-q/5}\, O\left(U_N(\omega)\right) \right\},
\end{equation*}
and (H\ref{hypo:densityderiv}) implies that 
\begin{equation} \label{eq:asymp_kde5}
\E \left( \int_I T_1(\omega,x) \, dx \right) =  O\left(N^{-4q/5} \right).
\end{equation}
(H\ref{hypo:densityderiv})-(H\ref{hypo:densityasym}) with (\ref{eq:asymp_kde3bis})-(\ref{eq:asymp_kde4}) results in
\begin{equation} \label{eq:asymp_kde6}
\E \left( \int_I T_2(\omega,x) \, dx \right) =  O\left(N^{-q} \right).
\end{equation}
With (\ref{eq:asymp_kde3a}),  (\ref{eq:asymp_kde5}), and (\ref{eq:asymp_kde6}),
\begin{equation} \label{eq:asymp_kde7}
\E \left( \int_I \var\left( \widehat{f}_t(x) | f_t  \right)\, dx \right) = O\left(N^{-4q/5} \right), 
\end{equation}
which matches the rate of convergence obtained for the bias term (see (\ref{eq:asymp_kde4})). Finally, (\ref{eq:asymp_kde1bis}), (\ref{eq:asymp_kde4}) and (\ref{eq:asymp_kde7}) entails
\begin{equation} \label{eq:asymp_kde8}
\E\| \widehat{f}_t - f_t \|^2 = O\left(N^{-4q/5} \right).
\end{equation}
Thanks to \eqref{eq:asymp_kde1} and \eqref{eq:asymp_kde8}, 
\begin{equation} \label{eq:asymp_kde9}
   \E\| \widehat{f}_t \ominus f_t \|_\mathcal{B}^2 = O\left(N^{-4q/5} \right), 
\end{equation}
which is the claimed result in $(i)$. 
\end{itemize}
{\em Proof of $(ii)$}. According to $(i)$, the rate of convergence $\delta_N$ involved in Theorem~\ref{theorem} is equal to $N^{-2q/5}$. By replacing $\delta_N$ with $N^{-2q/5}$ in the result of Theorem~\ref{theorem}, 
\begin{eqnarray*}
\left\| \widehat{m}_N(f) \ominus m(f) \right\|_\mathcal{B} \, = \, O(h_{\text{reg}}) + O\left(\left\{N^{4q/5} \, \pi_f(h_{\text{reg}})\right\}^{-1/2}\right) + O_P\left( \left\{N^{a/(1+a)} \, \pi_f(h_{\text{reg}})\right\}^{-1/2}\right). 
\end{eqnarray*}
The second term on the right side is negligible with respect to the third term as soon as $\displaystyle \frac{4\, q}{5}> \frac{a}{1+a}$ which is equivalent to the first condition $q > 5 a / \{ 4 ( 1 + a ) \}$ of (H\ref{hypo:densityasym}).

\section{Extension to time-dependent random vectors lengths \texorpdfstring{$n_t$}{nt} of  \texorpdfstring{$\bX_t$}{Xt}'s} \label{app:extension}

Instead of observing random vectors of same length $n$, it is possible to extend our framework to the setting of $n_t$-dimensional random vectors $\bX_t \coloneqq (X_{t,1},\ldots,X_{t,n_t})$, where the dimensions $n_t$ are time-dependent random integers. When $n_t$ depends on $N$ (i.e. $n_t = n_t(N)$), it is easy to define situations where our theoretical results still hold. For instance, set 
\begin{equation}\label{def:nt}
    n_t(N) \coloneqq \lfloor g(N) \times B_t \rfloor  \text{ a.s.},
\end{equation}
where $g(N) \rightarrow \infty$ as $N \rightarrow \infty$, and $\{B_t\}_{t\in\Z}$ is any random process (that may depend on $f_t$) such that, for each $t$, $B_t$ is a bounded rrv with $P(B_t > C)=1$ where $C>0$. This more complex probabilistic structure needs to precise the dependence structure between the joint process $\{(f_t, B_t)\}_{t\in\Z}$ and the process  $\{\bX_t\}_{t\in\Z}$ of observed $n_t(N)$-dimensional random vectors. Set $\mathcal{T}_p^q = \sigma\{(f_t,B_t): \, p\leq t \leq q\}$ and consider the quantity 
\begin{equation*}
\rho_{f,B}(H) = \sup\big\{ \left| \text{Corr}(U, \, V) \right|, \, U\in \mathcal{L}^2(\mathcal{T}_{-\infty}^0), \, V \in \mathcal{L}^2(\mathcal{T}_{H}^{+\infty})\big\}
\end{equation*}
which is the $\rho$-mixing coefficient of the joint process $\{(f_t,B_t)\}_{t\in\Z}$. Then, we replace (H\ref{hypo:dependence}) with:
\begin{itemize}
\item[(H\ref{hypo:dependence}')]
The random processes $\{f_t\}_{t\in \Z}$, $\{B_t\}_{t\in \Z}$ and $\{\bX_t\}_{t\in \Z}$ are such that:
\begin{itemize}
\item[$\bullet$] the joint process $\{(f_t, B_t)\}_{t\in \Z}$ is a stationary $\rho$-mixing process,
\item[$\bullet$] for any finite tuple $(t_1, \ldots, t_k) \in \Z^k$, the PDF of $(\bX_{t_1}, \ldots, \bX_{t_k})$ conditionally to $\{(f_s, Z_s)\}_{s\in \Z}$ is equal to $\Pi_{j=1}^k \, f_{t_j}^{n_{t_j}(N)}$.
\end{itemize}
\end{itemize}
\begin{rem}
    In the definition of $n_t(N)$ (see \eqref{def:nt}), it is crucial to separate the sources of randomness from the sample size $N$. This separation allows the definition of a dependence structure that does not involve $N$.
\end{rem}

\begin{description}

\item[\em Extension of Proposition~\ref{prop:dependence}.] To prove that Proposition~\ref{prop:dependence} still holds with (H\ref{hypo:dependence}'), just define the set $\mathcal{X} \coloneqq \cup_{k\geq 0}\left\{ (k, x): \, x\in I^k \right\}$ of disjoint unions with the convention $I^0 \coloneqq \{\emptyset\}$ where $I$ is the support of the PDFs. Although the length $n_t$ of $X_t$'s is a random integer, the random vectors $\bX_t$'s are valued in the same space $\mathcal{X}$. Then, with similar arguments as those used to proof Proposition~\ref{prop:dependence}, one can show that  
\begin{inparaenum}
\item[1)] under (H\ref{hypo:regression}), there exists a PDF-valued random process that satisfies the PDF-on-PDF nonparametric regression model, and the error process $\{\epsilon_t\}_{t\in \Z}$ is necessarily $\rho$-mixing, 
\item[2)] $\{\bX_t\}_{t\in \Z}$ is a stationary $\rho$-mixing process.
\end{inparaenum}

\item[\em Extension of Theorem~\ref{theorem}.] According to the new probabilistic and dependence structure, $\rho_f$ in (\ref{hypo:dependence}) is replaced with $\rho_{f,B}$, the $\rho$-mixing coefficient of the join process $\{ (f_t, B_t)\}_{t\in\Z}$. With the definition of $n_t(N)$, $n_t(N) \rightarrow \infty$ when $N \rightarrow \infty$ a.s. Then, one can still assume that, for each $t$, the mean integrated squared error of the estimator $\widehat{f}_t$ of $f_t$ depends on $N$ as in (H\ref{hypo:estdensity}):
\begin{equation}\label{hypo:MISE}
\E \left( \| \widehat{f}_t \ominus f_t \|^2_\mathcal{B} \right) = O(\delta_N^2).
\end{equation}
By adapting the proof of Theorem~\ref{theorem} to this new setting involving the join process $\{ (f_t, B_t)\}_{t\in\Z}$, and since $n_t(N)$ is only involved through the asymptotic behaviour of $\widehat{f}_t$ (see \eqref{hypo:MISE}),  Theorem~\ref{theorem} remains valid.

\item[\em Extension of Theorem~\ref{theo2:rate}.] The same rate of convergence for mean integrated squared error is obtained as soon as $g(N) \coloneqq N^q$ in \eqref{def:nt}. In this case, $n_t(N) \asymp N^q$ a.s., and Theorem~\ref{theo2:rate}-$(i)$ still holds; hence,  Theorem~\ref{theo2:rate} remains valid.

\end{description}

\bibliographystyle{agsm}
\bibliography{NFR_density.bib}

\end{document}